\documentclass[12pt]{article}
\usepackage{axodraw}
\usepackage{a4wide,epsfig,psfrag,amsmath,amssymb,cite,scalefnt,color}

\newcommand{\gsim}{\;\rlap{\lower 3.5 pt \hbox{$\mathchar \sim$}} \raise 1pt
 \hbox {$>$}\;}
\newcommand{\lsim}{\;\rlap{\lower 3.5 pt \hbox{$\mathchar \sim$}} \raise 1pt
 \hbox {$<$}\;}

\def\dblone{\hbox{$1\hskip -1.2pt\vrule depth 0pt height 1.6ex width 0.7pt
\vrule depth 0pt height 0.3pt width 0.12em$}}

\begin{document}


\title{\vskip-3cm{\baselineskip14pt
    \begin{flushleft}
      \normalsize SFB/CPP-11-45\\
      \normalsize TTP11-24
  \end{flushleft}}
  \vskip1.5cm
  Gluino-Squark Production at the LHC: The Threshold
}
\author{\small Matthias R. Kauth, Achim Kress and Johann H. K\"uhn\\[1em]
  {\small\it Institut f{\"u}r Theoretische Teilchenphysik}\\
  {\small\it Karlsruhe Institute of Technology (KIT)}\\
  {\small\it 76128 Karlsruhe, Germany}}

\date{}

\maketitle

\thispagestyle{empty}

\begin{abstract}

An analysis of the cross section for hadronic production of
gluino-squark pairs close to threshold is presented. Within the
framework of non-relativistic QCD a significant enhancement compared to
fixed order perturbation theory is observed which originates from the
characteristic remnants of the gluino-squark resonances below the
nominal pair threshold. The analysis includes all colour configurations
of $S$-wave gluino-squark pairs, i.e. triplet, sextet and ${\bf 15}$
representation. Matching coefficients at leading order are separately
evaluated for all colour configurations. The dominant QCD corrections,
arising from initial- and final-state radiation are included.  The
non-relativistic dynamics of the gluino pair is solved by calculating
the Green's function in Next-to-Leading Order (NLO). The results are
applied to benchmark scenarios, based on Snowmass Points and Slopes
(SPS). As a consequence of the large decay rate of at least one of the
constituents (squark or gluino), annihilation decays of the bound state
$\left(\tilde{g}\tilde{q}\right)\rightarrow gq$, $q\gamma$, $qZ$ or
$q'W^{\pm}$ are irrelevant. Thus the signatures of gluino-quark
production below and above the nominal threshold are
identical. Numerical results for the cross section at the Large Hadron
Collider (LHC) at $\sqrt{s}=7\,\mbox{TeV}$ and $14\,\mbox{TeV}$ are
presented. The enhancement of the total cross section through final
state interaction amounts to roughly $3\%$.

\noindent

\end{abstract}

\thispagestyle{empty}


\newpage

\section{\label{sec:intro}Introduction}
The search for new particles, predicted in supersymmetric models, is one
of the most important tasks of the experiments at the LHC. The detailed
determination of particle masses and couplings will be crucial for the
discrimination between various manifestations of supersymmetry (SUSY)
and alternative models, even more so if one wants to distinguish between
the different variants of supersymmetric models, to identify the origin
of breaking of supersymmetry and to measure the model parameters.  One
of the SUSY signals will be events with missing energy or missing
transverse momentum, resulting from cascade decays of squarks and
gluinos into the lightest supersymmetric particle (LSP) which escapes
detection. The existence of squarks, gluinos and the LSP is definitely a
key prediction of supersymmetry.
\par
Depending on details of the models, in particular the mass of the LSP,
gluinos and squarks with masses up to 3 TeV \cite{Baer:2009dn} could be
detected. The decay rates of gluinos and squarks depend strongly on the
ratio of squark and gluino masses. If the gluino is heavier ($m_{\tilde
g} > m_{\tilde q}$), the two-body decay of the gluino into a squark and
an antiquark (or its charge conjugate) is possible with the squark
decaying into a quark plus a chargino or neutralino. In the opposite
case ($m_{\tilde g} < m_{\tilde q}$) the two-body decay of the gluino is
kinematically forbidden and the now dominant three-body decay into
quark, antiquark and neutralino or chargino, mediated by the virtual
squark, leads to a small decay rate for the gluino while the squark
decays into the gluino plus a quark.
\par
The importance of squark and gluino searches has motivated a series of
detailed studies of hadroproduction cross sections for squarks and
gluinos. The lowest order has been evaluated long time ago
\cite{Harrison:1982yi,Haber:1984rc,Dawson:1983fw}. Subsequently NLO
SUSY-QCD corrections were calculated
\cite{Beenakker:1996ch,Beenakker:1996ed,Beenakker:1997ut}, more recently
the effect of soft-gluon resummation
\cite{Kulesza:2008jb,Beneke:2009rj,Langenfeld:2009eg,Kulesza:2009kq,
Beenakker:2009ha,Beenakker:2010nq,Beenakker:2011fu} was included. The
present paper will be concerned with gluino-squark-pair production close
to threshold, which exhibits a number of peculiar features.
\par
The sum of gluino and squark single decay width is expected to be much
larger than the annihilation decay width of the squark-gluino system. In
this case the bound state decay proceeds through the decay of the
constituents and for decay rates of several GeV no well defined bound
states exist. Nevertheless, final state interaction will lead to a
significant lowering of the effective production threshold, to an
enhancement of the cross section and a strong distortion of the
differential cross section, in particular of the distribution in the
invariant mass of the gluino-squark pair, with details depending on the
masses and single decay widths of gluinos and squarks.
\par
This scenario has many similarities with hadronic top quark-
\cite{Fadin:1990wx,Hagiwara:2008df,Kiyo:2008bv,Sumino:2010bv} and gluino-pair production
\cite{Hagiwara:2009hq,Kauth:2011vg} (with $m_{\tilde{q}}$ close to 
$m_{\tilde{g}}$)  close to threshold. 
(It differs, however, from gluino pair production in the case of
large squark masses, where annihilation decays of $(\tilde{g}\tilde{g})$
bound states into two gluon jets could still dominate above constituent
decays \cite{Keung:1983wz,Kuhn:1983sc,Goldman:1984mj,Kartvelishvili:1989pp,
Chikovani:1996bk,Kauth:2009ud,Kats:2009bv}.) In particular the distribution in the invariant mass of the gluino-squark
pair can be calculated with similar methods.
\par
A gluino-squark pair can be combined into bound states transforming
under three irreducible representations, two of them attractive and one
repulsive. As in Refs.~\cite{Hagiwara:2009hq,Hagiwara:2008df} the NLO
result for the hard corrections is approximated by including the
logarithmically enhanced terms from initial- and final-state
radiation. The Green's function is evaluated in NLO approximation.
\par
The paper will be organized as follows: For a self-contained treatment
we recall in Section\,\ref{sec:quantum} the quantum numbers of the bound
states, discuss various SUSY scenarios and present the qualitative
features of threshold production for the case of interest, i.e. for
gluino-squark bound states with decay rates comparable to the level
spacing of the would-be bound states.
\par
In Section\,\ref{subsec:Green} we will present the threshold enhancement
(or suppression) for the various colour configurations using Green's
functions in NLO approximation. These will be evaluated similar to those
of the $t\overline{t}$ system discussed in
Refs.~\cite{Hagiwara:2008df,Kiyo:2008bv} or the $\tilde{g}\tilde{g}$
system studied in \cite{Kauth:2011vg,Hagiwara:2009hq}. (For earlier
investigations of squark and gluino bound state production at hadron
colliders see
\cite{Keung:1983wz,Kuhn:1983sc,Goldman:1984mj,Kartvelishvili:1989pp,
Chikovani:1996bk,Kauth:2009ud,Kats:2009bv}.)  The important difference
is the non-degeneracy of the constituent masses, namely the gluino and
squark mass. In Section\,\ref{subsec:Short} the effect of initial- and
final-state radiation is investigated. Only $S$ waves will be
considered. The choice of the proper value of the strong coupling
$\alpha_{s}$ is discussed.
\par
Using this input, the hadronic production cross section can be evaluated
in a straightforward way in Section\,\ref{sec:num}. We limit the
discussion to proton-proton collisions at $7$ and $14$ TeV and give
results for several of the SUSY scenarios discussed in
Section\,\ref{sec:quantum} and compare the results to those obtained
without final-state corrections. Section\,\ref{sec:concl} contains our
conclusions.

\section{\label{sec:quantum}SUSY scenarios, gluino-squark bound states and threshold behaviour}
Let us briefly recall the quantum numbers of gluino-squark pairs in the
threshold region, classified according to their colour configurations
\cite{Kats:2009bv}. While the combination of a spin $1/2$ and a spin $0$
particle is trivial, a colour-octet and a colour-triplet state can be
combined into irreducible representations as follows (See also
Appendix~\ref{app:3times8}.):
\begin{eqnarray}
{\mathbf 3}\otimes{\mathbf 8}&=&{\mathbf 3}\oplus \overline{{\mathbf 6}}\oplus {\bf 15}
\,,
\label{eqn:3times8a}
\end{eqnarray}
while the same equation with a conjugate triplet on the left-hand side
(representing an antisquark) results in 
\begin{eqnarray}
\overline{{\mathbf 3}}\otimes {\mathbf 8}&=&\overline{{\mathbf 3}}\oplus{\mathbf 6}\oplus\overline{{\bf 15}}
\,.
\label{eqn:3times8b}
\end{eqnarray}
The interaction can be either attractive or repulsive. In lowest order
the coefficient of the QCD potential which governs the final state
interaction is given by the expectation value of the product of the
colour generators of the fundamental and the adjoint representation
$T^a_{ij}F^a_{kl}$. This product, in turn, can be expressed by
the eigenvalues of the quadratic Casimir operator of the constituents,
$C_F=4/3$ and $C_A=3$, and of the boundstate in representation $R$,
$C_R=(T^a+F^a)^2_R$:
\begin{eqnarray}
  T^{a,1}\cdot F^{a,2}&=&
  \frac{1}{2}\left[(T^{a,1}+F^{a,2})^2_R-(T^{a,1})^2-(F^{a,2})^2\right]=\frac{1}{2}\left(C_R-C_F-C_A\right)
  \,.
\label{eqn:FdotF}
\end{eqnarray}
The results are listed in Tab.~\ref{tab:table1}.
\begin{table}[tb]
\begin{center}
\begin{tabular}{c|ccc}
R & $\hspace{0.25cm}C_R\hspace{0.25cm}$ & $\hspace{0.25cm}T^{a,1}\cdot F^{a,2}\hspace{0.25cm}$ & $\hspace{0.25cm}\mbox{interaction}\hspace{0.25cm}$ \\ \hline
${\mathbf3},\overline{{\mathbf3}}$ & $\frac{4}{3}$ & $-\frac{3}{2}$ & \mbox{attractive}\\
${\mathbf6},\overline{{\mathbf6}}$ & $\frac{10}{3}$ & $-\frac{1}{2}$ & \mbox{attractive}\\
$\hspace{0.25cm}{\bf15},\overline{{\bf15}}\hspace{0.25cm}$ & $\frac{16}{3}$ & $\frac{1}{2}$ & \mbox{repulsive}
\end{tabular}
\caption{Colour interaction of a $SU(3)$ octet and triplet
  respectively antitriplet.}
\label{tab:table1}
\end{center}
\end{table}
\par
The interaction potential between gluino and squark in lowest order is
then given by the ``Coulomb'' potential and depends on the distance
$r\equiv\left|\mathbf{r}\right|$ between the two constituents
\begin{eqnarray}
V^{[R]}_{C,\mbox{\tiny Born}}\left(\mathbf{r}\right)&=&-\frac{C^{[R]}\,\alpha_s}{r}\,,
\label{eqn:Vcoul}
\end{eqnarray}
with
\begin{eqnarray}
C^{[R]}&=&C^{[\overline{R}]}\hspace{0.2cm}=\hspace{0.2cm}\frac{3}{2},\,\frac{1}{2},\,-\frac{1}{2}\hspace{0.5cm}\mbox{for}\hspace{0.5cm}R\hspace{0.2cm}=\hspace{0.2cm}{\mathbf3},\,{\mathbf6},\,{\bf15}
\,.
\label{eqn:Vcoul2}
\end{eqnarray}
For the cases with negative (positive) coefficients, corresponding to
attraction (repulsion), the cross section will be enhanced
(suppressed). NLO corrections are discussed in the next chapter.
\par
The amplitudes for the production and decay of gluino and squark have to
be disentangled according to the corresponding representation. It is
convenient to define the projectors
\begin{eqnarray}
\mathbb{P}^{[R]}_{ai,bj}&\equiv&a^{[R]}\,\delta_{ab}\,\dblone_{ij}+b^{[R]}\,d_{abm}\,T^m_{ij}+c^{[R]}\,i\,f_{abm}\,T^m_{ij}
\,,
\label{eqn:PROJ}
\end{eqnarray}
with the coefficients listed in Tab.~\ref{tab:PROJ} which obey
\begin{eqnarray}
\sum_i\mathbb{P}^{[R_i]}_{aj,bk}&=&\delta_{ab}\,\dblone_{jk}\,,
\nonumber\\
\mathbb{P}^{[R_1]}_{ai,bj}\mathbb{P}^{[R_2]}_{bj,ck}&=&\mathbb{P}^{[R_1]}_{ai,ck}\,\delta_{R_1R_2}\,,
\nonumber\\
\mathbb{P}^{[R]}_{ai,ai}&=&R
\,.
\label{eqn:PROJ2}
\end{eqnarray}
The corresponding projectors for a general $SU(N)$ and a sketch of their
derivation are given in Appendix~\ref{app:3times8}, together with the
results for the coefficient of the potential, $C^{[R]}$.
\begin{table}[tb]
\begin{center}
\begin{tabular}{c||c|c|c}
$R$ & $\hspace{0.2cm}a^{[R]}\hspace{0.2cm}$ & $\hspace{0.2cm}b^{[R]}\hspace{0.2cm}$ & $\hspace{0.2cm}c^{[R]}\hspace{0.2cm}$ \\\hline
${\mathbf3},\overline{\mathbf{3}}$ & $\frac{1}{8}$ & $\frac{3}{8}$ & $\frac{3}{8}$\\
${\mathbf6},\overline{\mathbf{6}}$ & $\frac{1}{4}$ & $-\frac{3}{4}$ & $\frac{1}{4}$\\
$\hspace{0.2cm}{\bf15},\overline{{\bf15}}\hspace{0.2cm}$ & $\frac{5}{8}$ & $\frac{3}{8}$ & $-\frac{5}{8}$
\end{tabular}
\caption{Coefficients of the projectors from Eq.~(\ref{eqn:PROJ})
  within $SU(3)$.}
\label{tab:PROJ}
\end{center}
\end{table}
\par
Restricting ourselves to the threshold region, only S-wave
configurations will be retained. The cross section of states with higher
angular momenta is suppressed by at least two powers of the relative
velocity $v$ of the two constituents. The classification described in
Tab.~\ref{tab:table1} is applicable to bound state and continuum
production and will be important for the description of final state
interaction.
\par
As mentioned in the Introduction the phenomenology of bound state
production is governed by the relative size of the sum of the decay
rates of constituents, i.e. of gluino and squark,
$\Gamma_{\tilde{g}}+\Gamma_{\tilde{q}}$, compared to the the level
spacing $\Delta M$ between the ground state and the first radial
excitation of the strongest bound colour triplet configuration. The
choice of $\Delta M$ is motivated by the fact that the binding energy
per se depends evidently on the choice of the mass definition (pole
mass, potential subtracted mass, \ldots) while $\Delta M$ is convention
independent. The full decay rate of the gluino-squark system is given
as sum of the two single decay rates plus the annihilation decay
rate.
\par
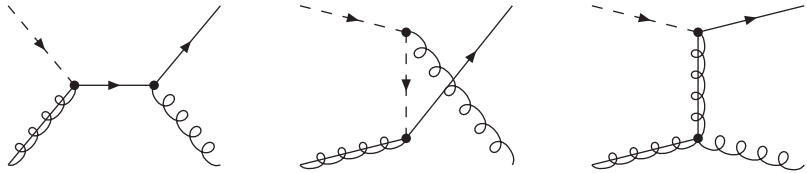
\begin{figure}[tb]
\begin{center}
\begin{picture}(300,60)(0,0)
\SetColor{Black}
\DashArrowLine(0,60)(25,30){4}
\Vertex(25,30){1.8}
\ArrowLine(25,30)(55,30)
\Vertex(55,30){1.8}
\ArrowLine(55,30)(80,60)
\Gluon(55,30)(80,0){-3}{4}
\Gluon(0,0)(25,30){-2.5}{4}
\Line(0,0)(25,30)
\Vertex(150,50){1.8}
\DashArrowLine(150,50)(150,10){4}
\Vertex(150,10){1.8}
\DashArrowLine(110,60)(150,50){4}
\Gluon(110,0)(150,10){-2.5}{4}
\Line(110,0)(150,10)
\Gluon(150,50)(190,0){3}{6}
\Line(150,10)(160,22.5)
\ArrowLine(160,22.5)(190,60)
\Vertex(260,50){1.8}
\Line(260,50)(260,10)
\Gluon(260,50)(260,10){2.5}{4}
\Vertex(260,10){1.8}
\DashArrowLine(220,60)(260,50){4}
\Gluon(220,0)(260,10){-2.5}{4}
\Line(220,0)(260,10)
\Gluon(260,10)(300,0){-3}{4}
\ArrowLine(260,50)(300,60)
\end{picture}
\caption{Feynman diagrams contributing at LO to
  $\tilde{g}\tilde{q}\rightarrow gq$.}
\label{fig:DecayStrong}
\end{center}
\end{figure}
The annihilation of the gluino-squark system is dominated by the strong
decay into a gluon and a quark. The corresponding Feynman diagrams are
given in Fig.~\ref{fig:DecayStrong}. We assume massless quarks and
``Minimal Flavour Violation'' (MFV) where squarks couple only the quarks
of the same flavour. Let us in a first step evaluate the amplitude for
$\tilde{g}\tilde{q}$ production by a gluon and a quark, a result also
required for the gluino-squark annihilation. At threshold (i.e. for
$p_{\tilde{g}}=m_{\tilde{g}}/(m_{\tilde{g}}+m_{\tilde{q}})P$ and
$p_{\tilde{q}}=m_{\tilde{q}}/(m_{\tilde{g}}+m_{\tilde{q}})P$) the
amplitude $\mathcal{M}$ can be decomposed into the contributions from
three irreducible representations
\begin{subequations}
\begin{eqnarray}
\mathcal{M}\left(gq\rightarrow\tilde{g}\tilde{q}_i\right)&=&
i\sqrt{2}\,g^2\varepsilon_{\mu}\left(p_g\right)\overline{u}\left(p_{\tilde{g}},m_{\tilde{g}}\right)
\nonumber\\
&&\times\biggl\{
\,\frac{1}{s}\left(m_{\tilde{q}_i}+m_{\tilde{g}}\right)\gamma^{\mu}\,T_{jk}^bT_{kl}^a
\nonumber\\
&&\hspace{0.75cm}+\frac{1}{t-m_{\tilde{g}}^2}\,\gamma^{\mu}\left(m_{\tilde{g}}-p_{\tilde{q}_i}\hspace{-0.44cm}/\hspace{0.15cm}\right)\,if_{bac}T_{jl}^c
\nonumber\\
&&\hspace{0.75cm}+\frac{2p_{\tilde{q}_i}^{\mu}}{u-m_{\tilde{q}_i}^2}\,T_{jk}^aT_{kl}^b
\biggr\}\left(P_RU_{i,R}-P_LU_{i,L}\right)u\,(p_q,0)
\,,
\label{eqn:amp}
\\
&=&-i\,\frac{\sqrt{2}\,g^2}{6(m_{\tilde{g}}+m_{\tilde{q}_i})m_{\tilde{g}}}\,\overline{u}\left(p_{\tilde{g}},m_{\tilde{g}}\right)\varepsilon\hspace{-0.18cm}/\hspace{0.0cm}(p_g)\left(P_RU_{i,R}-P_LU_{i,L}\right)u\,(p_q,0)
\nonumber\\
&&\times\Bigl[
\left(m_{\tilde{g}}+9m_{\tilde{q}_i}\right)\mathbb{P}^{[{\bf 3}]}_{bj,al}
+3\left(m_{\tilde{g}}+m_{\tilde{q}_i}\right)\mathbb{P}^{[{\bf \overline{6}}]}_{bj,al}
-3\left(m_{\tilde{g}}+m_{\tilde{q}_i}\right)\mathbb{P}^{[{\bf 15}]}_{bj,al}
\Bigr]\,.
\nonumber\\
\label{eqn:amp2}
\end{eqnarray}
\end{subequations}
(Here we have used $p_g\cdot\varepsilon(p_g)=P\cdot\varepsilon(p_g)=0$.)
with the Mandelstam variables $s$, $t$, $u$ and the elements of the
orthogonal squark-mixing matrix
\begin{eqnarray}
U&=&\left(\begin{array}{cc}U_{LL}&U_{LR}\\U_{RL}&U_{RR}\end{array}\right)
\,.
\label{eqn:squmix}
\end{eqnarray}
The index $i$ in Eqs.~(\ref{eqn:amp}), (\ref{eqn:amp2}) labels the chiralities $R$ and $L$.
For the decay of the triplet and sextet ground state we find (in
agreement\footnote{We thank Y. Kats and D. Kahawala for correspondence on this point.} 
with \cite{Kahawala:2011pc})
\begin{eqnarray}
\Gamma\bigl((\tilde{g}\tilde{q})_{{\bf 3}/{\bf \overline{3}}}\rightarrow gq\bigr)&=&
\frac{\alpha_{s}^2\bigl|R^{[{\bf 3}]}_1(0)\bigr|^2(m_{\tilde{g}}+9m_{\tilde{q}})^2}%
{144\,m_{\tilde{g}}^2\,m_{\tilde{q}} (m_{\tilde{g}}+m_{\tilde{q}}) }\,,
\nonumber\\
\Gamma\bigl((\tilde{g}\tilde{q})_{{\bf 6}/{\bf \overline{6}}}\rightarrow gq\bigr)&=&
\frac{\alpha_{s}^2\bigl|R^{[{\bf 6}]}_1(0)\bigr|^2(m_{\tilde{g}}+m_{\tilde{q}})}{16\,m_{\tilde{g}}^2\,m_{\tilde{q}}}
\,.
\end{eqnarray}
The elements of the squark-mixing matrix drop out after summation over
quark helicities and in the limit of massless quarks.
\par
\begin{figure}[tb]
\begin{center}
\begin{picture}(190,60)(20,0)
\SetColor{Black}
\DashArrowLine(0,60)(25,30){4}
\Vertex(25,30){1.8}
\ArrowLine(25,30)(55,30)
\Vertex(55,30){1.8}
\ArrowLine(55,30)(80,0)
\Photon(55,30)(80,60){-3}{3}
\Gluon(0,0)(25,30){-2.5}{4}
\Line(0,0)(25,30)
\Vertex(150,50){1.8}
\DashArrowLine(150,50)(150,10){4}
\Vertex(150,10){1.8}
\DashArrowLine(110,60)(150,50){4}
\Gluon(110,0)(150,10){-2.5}{4}
\Line(110,0)(150,10)
\Photon(150,50)(190,60){3}{3}
\ArrowLine(150,10)(190,0)
\Text(195,60)[l]{$\gamma, Z, W^{\pm}$}
\end{picture}
\caption{Feynman diagrams contributing at LO to
  $\tilde{g}\tilde{q}\rightarrow \gamma q,Zq,W^{\pm}q'$.}
\label{fig:feyn}
\end{center}
\end{figure}
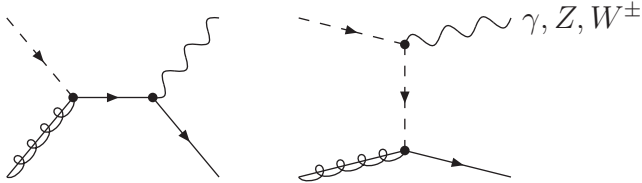

The radial part of the Schr\"odinger wave function at the origin can
be derived from the one of the quarkonia or gluinonia by replacing the
colour factor and the reduced mass
\begin{eqnarray}
\left|R^{[{\bf 3}/{\bf 6}]}_n(0)\right|^2&=&\frac{4\left(C^{[{\bf 3}/{\bf 6}]}\,\alpha_s\right)^3m_{\rm red}^3}{n^3}
\,,
\end{eqnarray}
with $m_{\rm
red}=m_{\tilde{g}}m_{\tilde{q}}/(m_{\tilde{g}}+m_{\tilde{q}})$. Numerically,
for $m_{\tilde{g}}\sim m_{\tilde{q}}\sim m\sim1\,\mbox{TeV}$, this leads
to a decay rate $\Gamma=(75/32)\,\alpha_s^2\alpha_s^3\,m$ which is of
$\mathcal{O}\left(10\,\mbox{MeV}\right)$ and thus dramatically smaller
than the constituent decay rate discussed below. Gluino-squark
annihilation into quarks and electroweak gauge bosons
(Fig.~\ref{fig:feyn}) is further suppressed by the electromagnetic or
weak coupling and thus completely negligible.
\par
The threshold behaviour, i.e. the dependence of the cross section on the
invariant mass of the gluino-squark system is governed by the relative
size of the constituent decay rate and the binding energy, or more
precisely, the excitation energy of the dominant ground
state\footnote{Final state interaction for $\tilde{g}\tilde{q}$ in the
sextet representation is suppressed by powers of $C^{[{\bf 6}]}/C^{[{\bf
3}]}=1/3$.} $\Delta M\equiv E_{2}^{[{\bf 3}]}-E_{1}^{[{\bf 3}]}$ with
\begin{eqnarray}
E^{[{\bf 3}]}_n&=&-\frac{\left(C^{[{\bf 3}]}\,\alpha_s\right)^2\,m_{\rm red}}{2\,n^2}
\,.
\label{eqn:E3}
\end{eqnarray}
(Higher order corrections are included in our evaluation of the Green's
function in Section~\ref{subsec:Green}.)
\par
\par
\begin{table}
\begin{center}
\begin{tabular}{c||c|c|c|c|c|c|c}
$\hspace{0.05cm}\mbox{benchmark}\hspace{0.05cm}$&$\,m_{\tilde{g}}\,$&$\,\overline{m}_{\tilde{q}}\,$&$\hspace{0.15cm}\Gamma_{\tilde{g}}+\Gamma_{\tilde{u}_L}\hspace{0.15cm}$&$\hspace{0.15cm}\Gamma_{\tilde{g}}+\Gamma_{\tilde{u}_R}\hspace{0.15cm}$&$\hspace{0.15cm}\Gamma_{\tilde{g}}+\Gamma_{\tilde{d}_L}\hspace{0.15cm}$&$\hspace{0.15cm}\Gamma_{\tilde{g}}+\Gamma_{\tilde{d}_R}\hspace{0.15cm}$&$\,\Delta
\overline{M}_{\tilde{g}\tilde{q}}\,$
\\
point&$\hspace{0.4cm}\left[\mbox{GeV}\right]\hspace{0.4cm}$&$\hspace{0.4cm}\left[\mbox{GeV}\right]\hspace{0.4cm}$&$\left[\mbox{GeV}\right]$&$\left[\mbox{GeV}\right]$&$\left[\mbox{GeV}\right]$&$\left[\mbox{GeV}\right]$&$\hspace{0.3cm}\left[\mbox{GeV}\right]\hspace{0.3cm}$
\\\hline\hline
(a)&$606.11$&$555.39$&$10.00$&$5.69$&$9.80$&$4.83$&$3.28$
\\\hline
(b)&$493.05$&$453.62$&$7.65$&$4.07$&$7.40$&$3.38$&$2.82$
\\\hline
(c)&$381.45$&$349.95$&$6.12$&$3.31$&$5.80$&$2.78$&$2.34$
\\\hline
(d)&$717.12$&$655.32$&$12.37$&$7.35$&$12.21$&$6.34$&$3.72$
\\\hline
(e)&$826.71$&$753.94$&$14.73$&$9.02$&$14.58$&$7.85$&$4.13$
\\\hline
(f)&$935.18$&$851.53$&$17.05$&$10.68$&$16.92$&$9.36$&$4.53$
\\\hline
(g)&$1042.60$&$948.11$&$19.33$&$12.31$&$19.22$&$10.84$&$4.91$
\\\hline
(h)&$1149.42$&$1044.12$&$21.57$&$13.91$&$21.47$&$12.29$&$5.29$
\\\hline
(i)&$936.42$&$859.67$&$16.42$&$9.97$&$16.28$&$8.63$&$4.55$
\\\hline
(j)&$802.21$&$1554.00$&$79.95$&$65.25$&$80.04$&$62.71$&$5.16$
\\\hline
(k)&$566.65$&$1299.75$&$79.29$&$67.00$&$79.40$&$64.93$&$4.13$
\\\hline
(l)&$319.59$&$1055.75$&$79.34$&$69.31$&$79.45$&$67.68$&$2.90$
\\\hline
(m)&$1030.98$&$1812.80$&$81.51$&$64.35$&$81.59$&$61.35$&$6.08$
\\\hline
(n)&$1255.61$&$2073.80$&$83.80$&$64.20$&$83.87$&$60.71$&$6.95$
\\\hline
(o)&$933.03$&$841.02$&$17.63$&$11.36$&$17.50$&$10.06$&$4.50$
\\\hline
(p)&$734.11$&$753.47$&$9.59$&$3.34$&$9.52$&$2.17$&$3.95$
\\\hline
(q)&$719.66$&$664.46$&$17.96$&$12.85$&$17.87$&$11.82$&$3.74$
\end{tabular}
\caption{Comparison of the gluino mass, the averaged squark mass of the
  first generation, the sum of the single decay widths and the level
  spacing for 26 benchmark points defined in Tab.~4 of
  Ref.~\cite{Kauth:2011vg}.}
\label{tab:SPS}
\end{center}
\end{table}
In Tab.~\ref{tab:SPS} the sum of the single decay rates, calculated with
the program {\tt SDECAY} \cite{Muhlleitner:2003vg}, is compared to the
level spacing $\Delta M$ for the 17 benchmark points defined in Tab.~4
of Ref.~\cite{Kauth:2011vg} (the spectrum of the SUSY masses has been
derived with {\tt SuSpect} \cite{Djouadi:2002ze}). Due to MFV and the
structure of the proton only the gluino-squark systems of the first
generation are of interest. For the level spacing
$\Delta\overline{M}_{\tilde{g}\tilde{q}}$ the averaged squark mass of
the first generation
\begin{eqnarray}
\overline{m}_{\tilde{q}}&\equiv&\frac{m_{\tilde{u}_L}+m_{\tilde{u}_R}+m_{\tilde{d}_L}+m_{\tilde{d}_R}}{4}
\,,
\label{eqn:Msqu1}
\end{eqnarray}
has been used. Obviously the sum of the single decay rates is of order
$\Delta\overline{M}_{\tilde{g}\tilde{q}}$ (benchmark points (a)-(i) and
(o)-(q)) or far larger (benchmark points (j)-(n)). In the latter cases
the width of the gluino is small but the one of the squark is huge in
comparison with the mass splitting $\Delta\overline{M}_{\tilde{g}\tilde{q}}$.
\par
The annihilation decay rate is always negligible and thus not shown
here. Qualitatively the dominance of constituent decays is easily
understood: For squark masses significantly larger than gluino masses
the squark decay rate with its kinematically enhanced two-body mode
becomes large. In the opposite case the gluino decay rate
increases. Only in the degenerate case $m_{\tilde{g}}\sim m_{\tilde{q}}$
both rates are relatively small, i.e. comparable to $\Delta M$, and it
is only in this case that resonant structures will arise. Nevertheless,
even in the case where the decay rate is significantly larger than
$\Delta M$, final state interaction will lead to a broad
enhancement. Therefore results for two benchmark points will be
presented in Section~\ref{sec:num}, one with
$\Gamma_{\tilde{g}}+\Gamma_{\tilde{q}}$ comparable to $\Delta M$, the
second one with $\Gamma_{\tilde{g}}+\Gamma_{\tilde{q}}$ significantly
larger.
\par
Hence formation of bound states \cite{Kauth:2009ud} or of sharp
resonances in the differential cross section (as discussed for the
gluino-pair production in \cite{Hagiwara:2009hq,Kauth:2011vg}) is not
possible here. Nevertheless, for the benchmark points (a)-(i) and
(o)-(q) the binding effects of the colour triplet and octet will be
visible in the differential cross section.
\par
The technical aspects of the calculation for threshold production are
very similar to the ones for top-quarks
\cite{Hagiwara:2008df,Kiyo:2008bv} and gluinos
\cite{Hagiwara:2009hq,Kauth:2011vg}. The cross section, differential in
$M$, the invariant mass of the gluino-squark pair, can be decomposed
into a factor representing the hard, short distance part and a factor
determined by the imaginary part of the Green's function, evaluated at
the origin. This partonic cross section is then convoluted with the
luminosity function:
\begin{eqnarray}
M\frac{\mbox{d}\sigma_{PP\rightarrow
    R}}{\mbox{d}M}(S,M^2)&=&\sum_{i,j}\int_{\rho}^1\mbox{d}\tau\left[\frac{\mbox{d}\mathcal{L}_{ij}}{\mbox{d}\tau}\right]\left(\tau,\mu_F^2\right)\,M\frac{\mbox{d}\hat{\sigma}_{ij\rightarrow
    R}}{\mbox{d}M}(\hat{s},M^2,\mu_R^2,\mu_F^2)
\,,
\label{eqn:masterSG}
\end{eqnarray}
with
\begin{eqnarray}
M\frac{\mbox{d}\hat{\sigma}_{ij\rightarrow
    R}}{\mbox{d}M}(\hat{s},M^2,\mu_R^2,\mu_F^2)&\hspace{-0.cm}=&\hspace{-0.cm}\mathcal{F}_{ij\rightarrow
  R}(\hat{s},M^2,\mu_R^2,\mu_F^2)\,\frac{4\,\mbox{Im}\left\{G^{[R]}\left(\mathbf{0},M+\frac{i}{2}\left(\Gamma_{\tilde{g}}+\Gamma_{\tilde{q}_i}\right)\right)\right\}}{m_{\rm red}^2}\,,
\nonumber\\
\left[\frac{\mbox{d}\mathcal{L}_{ij}}{\mbox{d}\tau}\right]\left(\tau,\mu_F^2\right)&=&\int_0^1\mbox{d}x_1\int_0^1\mbox{d}x_2\,f_{i|P}(x_1,\mu_F^2)\,f_{j|P}(x_2,\mu_F^2)\,\delta\left(\tau-x_1x_2\right)
\,.
\label{eqn:masterSG2}
\end{eqnarray}
The quantities $S$ and $\hat{s}$ denote as usual the hadronic and the
partonic center of mass energy squared, and $\tau\equiv\hat{s}/S$. The
lower limit of the integration is given by $\rho\equiv M^2/S$. The
superscript of the Green's function refers to the colour representation
$R$ of the corresponding $S$ wave and $\mu_F$ and $\mu_R$ denote the
factorization and renormalization scale, respectively.
\par
The only reaction present at leading order (LO) is
$gq\rightarrow\tilde{g}\tilde{q}_i$ where in the Feynman diagrams of
Fig.~\ref{fig:DecayStrong} initial and final states have to be
exchanged. For the four squark types under consideration and including
both squark and antisquark the hard parts of the cross sections are
given as
\begin{eqnarray}
\mathcal{F}_{ij\rightarrow R}^{(0)}&=&\mathcal{N}_{ij}^{[R]}\frac{\pi^2\alpha_s^2(\mu_R)}{12\hat{s}}\delta(1-z)
\,,
\label{eqn:FloSG}
\end{eqnarray}
with $z=M^2/\hat{s}$. The non-vanishing normalization factors
$\mathcal{N}_{ij}^{[R]}$ are obtained from Eqs.~(\ref{eqn:amp}) and
(\ref{eqn:amp2}):
\begin{eqnarray}
\mathcal{N}_{gq}^{[{\bf
    3}]}&=&\mathcal{N}_{g\overline{q}}^{[\overline{\bf 3}]}\hspace{0.2cm}=\hspace{0.2cm}\frac{(m_{\tilde{g}}+9\overline{m}_{\tilde{q}})^2}{3m_{\tilde{g}}(m_{\tilde{g}}+\overline{m}_{\tilde{q}})}\,,
\nonumber\\
\mathcal{N}_{gq}^{[\overline{\bf
    6}]}&=&\mathcal{N}_{g\overline{q}}^{[{\bf 6}]}\hspace{0.2cm}=\hspace{0.2cm}\frac{6(m_{\tilde{g}}+\overline{m}_{\tilde{q}})}{m_{\tilde{g}}}\,,
\nonumber\\
\mathcal{N}_{gq}^{[{\bf
    15}]}&=&\mathcal{N}_{g\overline{q}}^{[\overline{\bf 15}]}\hspace{0.2cm}=\hspace{0.2cm}\frac{15(m_{\tilde{g}}+\overline{m}_{\tilde{q}})}{m_{\tilde{g}}}
\,,
\label{eqn:NloSG}
\end{eqnarray}
which do not depend on the squark mixing.
\par
The Green's function depends on the energy
$E=M-(m_{\tilde{g}}+m_{\tilde{q}_i})$ and on the sum of the decay
rates. It is obtained from the non-relativistic Schr\"odinger equation
\begin{eqnarray}
\left\{\left[\frac{2\left(-i\mathbf{\nabla}\right)^2}{m_{\rm
        red}}+V^{[R]}(\mathbf{r})\right]-\left(E+\frac{i}{2}\left(\Gamma_{\tilde{g}}+\Gamma_{\tilde{q}_i}\right)\right)\right\}\,G^{[R]}\left(\mathbf{r},M+\frac{i}{2}\left(\Gamma_{\tilde{g}}+\Gamma_{\tilde{q}_i}\right)\right)&\hspace{-0.2cm}=&\hspace{-0.2cm}\delta^{(3)}\left(\mathbf{r}\right)
\,,
\nonumber\\
\label{eqn:SGL}
\end{eqnarray}
with the potential being $V_{C,\mbox{\tiny Born}}^{[R]}(\mathbf{r})$,
the Coulomb potential of Eqs.~(\ref{eqn:Vcoul}) and
(\ref{eqn:Vcoul2}). NLO corrections to both, the hard parts of the cross
section as far as the Green's function are studied in the next section.
\section{\label{sec:NLO}Next to leading order corrections}
The NLO corrections of the cross section can be split up into those for
the imaginary part of the Green's function
$\mbox{Im}\bigl\{G^{[R]}\bigr\}$ and those for the hard coefficients
$\mathcal{F}_{ij\rightarrow R}$ as defined in Eq.~(\ref{eqn:masterSG2}).
\subsection{\label{subsec:Green}Green's function}
Following the idea of the Green's function method developed in
Refs.~\cite{Fadin:1987wz,Fadin:1988fn} we start with the interaction
potential in position space. With the colour coefficients of
Eq.~(\ref{eqn:Vcoul2}) it is given up to NLO
\begin{eqnarray}
V_C^{[R]}\left(\bf{r}\right)&=&-C^{[R]}\frac{\alpha_{s}\bigl(\mu_G^{[R]}\bigr)}{r}\left\{1+\frac{\alpha_{s}\bigl(\mu_G^{[R]}\bigr)}{4\pi}\biggl[2\beta_{0}\left(\ln\left(\mu_G^{[R]}r\right)+\gamma_E\right)+a_{1}\biggr]\right\}
\,,
\label{eqn:potNLOsg}
\end{eqnarray}
The coefficient $a_1=31/9C_A-20/9T_Fn_f$ is identical to the one
relevant for the NLO corrections to the potentials of gluinonia and
quarkonia \cite{Collet:2011kq}.  The coefficient of the QCD-beta
function is $\beta_0=11/3C_A-4/3T_Fn_f$ with $C_A=3$ and $T_F=1/2$ and
$\gamma_E=0.5772$. The choice of $n_f=5$ active quark flavours is
motivated by the scale of the Green's function (see Eq.~(\ref{eqn:muG}))
for the gluino and squark masses under consideration. Effects of massive
top quarks could be incorporated, however, are irrelevant at the present
level of precision. For both LO and NLO we adopt
$\alpha_s(M_Z=91.1876\,\mbox{GeV})=0.12018$ as provided from MSTW2008NLO
\cite{Martin:2009iq} and employ the two-loop running as provided by {\tt
RunDec} \cite{Chetyrkin:2000yt}.
\par
The Green's function for the top-antitop system is known in compact
analytic form \cite{Beneke:1999qg} (see also \cite{Pineda:2006ri}) and
the result is easily applied to the present case
\begin{eqnarray}
G^{[R]}\left(E+\frac{i}{2}\left(\Gamma_{\tilde{g}}+\Gamma_{\tilde{q}_i}\right)\right)&\hspace{-0.05cm}\equiv&\hspace{-0.05cm}G^{[R]}\left(\mathbf{0},E+m_{\tilde{g}}+m_{\tilde{q}_i}+\frac{i}{2}\left(\Gamma_{\tilde{g}}+\Gamma_{\tilde{q}_i}\right)\right)
\nonumber\\
&=&\hspace{-0.05cm}i\,\frac{v\,m_{\rm
    red}^2}{\pi}+\frac{C^{[R]}\,\alpha_{s}\bigl(\mu_G^{[R]}\bigr)m_{\rm
    red}^2}{\pi}\left[g_{\mbox{\scriptsize
      LO}}+\frac{\alpha_{s}\bigl(\mu_G^{[R]}\bigr)}{4\pi}g_{\mbox{\scriptsize
      NLO}}+\,\ldots\,\right]\,,
\nonumber\\
\label{eqn:Green}
\end{eqnarray}
with
\begin{eqnarray}
g_{\mbox{\scriptsize LO}}&\equiv&L-\psi^{(0)}\,,
\nonumber\\
g_{\mbox{\scriptsize NLO}}&\equiv&\beta_{0}\biggl[\,L^2-2L\left(\psi^{(0)}-\kappa\psi^{(1)}\right)+\kappa\psi^{(2)}+\left(\psi^{(0)}\right)^2-3\psi^{(1)}-2\kappa\psi^{(0)}\psi^{(1)}
\nonumber\\
&&\hspace{0.65cm}+4\,_{4}F_{3}\left(1,1,1,1;2,2,1-\kappa;1\right)\biggr]+a_{1}\biggl[L-\psi^{(0)}+\kappa\psi^{(1)}\biggr]
\,,
\label{eqn:Green2}
\end{eqnarray}
and
\begin{eqnarray}
\kappa\hspace{0.2cm}\equiv\hspace{0.2cm}i\frac{C^{[R]}\,\alpha_{s}\bigl(\mu_G^{[R]}\bigr)}{2\,v}\,,\hspace{0.5cm}
v\hspace{0.2cm}\equiv\hspace{0.2cm}\sqrt{\frac{E+\frac{i}{2}\left(\Gamma_{\tilde{g}}+\Gamma_{\tilde{q}_i}\right)}{2m_{\rm red}}}\,,\hspace{0.5cm}
L\hspace{0.2cm}\equiv\hspace{0.2cm}\ln\frac{i\,\mu_G^{[R]}}{4\,m_{\rm red}\,v}
\,.
\label{eqn:Green3}
\end{eqnarray}
The $n$-th derivative $\psi^{(n)}=\psi^{(n)}(1-\kappa)$ of the digamma
function $\psi(z)=\gamma_{E}+(d/dz)\ln\Gamma(z)$ is evaluated at
$(1-\kappa)$. The numerical solution of the Schr\"odinger equation of
Eq.~(\ref{eqn:SGL}) involves multiple poles in the binding energy which
have to be resummed to single poles \cite{Beneke:1999qg}.  The
definition of the Generalized Hypergeometric Function $_4F_3$ together
with comments about its numerical evaluation can be found in
Ref.~\cite{Kauth:2009ud}. In particular, attention has to be paid to the
evaluation of $_4F_3$ at the beginning of a branch cut for its last
argument.
\subsection{\label{subsec:Short}Short distance corrections}
The complete NLO corrections to the hard part of the cross section
involve virtual corrections to the process
$gq\rightarrow\tilde{g}\tilde{q}$ plus real radiation (for 
gluino-pair production see \cite{Kauth:2011vg}). The production of the
final state $\tilde{g}\tilde{u}_1$ in colour representation $R$, for
example, will receive contributions from the processes $gg\rightarrow
R\overline{u}$, $ug\rightarrow Rg$, $uu\rightarrow Ru$,
$u\overline{u}\rightarrow R\overline{u}$, $uq\rightarrow Rq$,
$u\overline{q}\rightarrow R\overline{q}$ and $q\overline{q}\rightarrow
R\overline{q}$ with $q\neq u$.
\par
For the present analysis the complete calculation has been replaced by
an approximation employed in \cite{Hagiwara:2008df,Hagiwara:2009hq} for
$t\overline{t}$ and $\tilde{g}\tilde{g}$ production which includes the
leading logarithms from initial and final state radiation, affecting the
subprocess $gq\rightarrow\tilde{g}\tilde{q}$ which was present at LO
already. The quality of this approximation can be judged e.\ g. from the
corresponding analysis for top production
(\cite{Kiyo:2008bv}, Eq.~(14)) where both partial and
complete results are available. In this case the size of the corrections
depends on the subprocess. For gluon induced processes the corrections are
large (35\% -- 40\% of the corrected result) with the enhanced term
amounting to about 3/4 of the corrections. For quark induced processes,
however, the corrections amount to  25\% -- 30\% with roughly equal
share of enhanced and non-enhanced terms. One gets
\begin{eqnarray}
\mathcal{F}_{ij\rightarrow
  R}&=&\mathcal{N}_{ij}^{[R]}\,\frac{\pi^2\,\alpha_s^2(\mu_R)}{12\,\hat{s}}\left[1+\frac{\alpha_s(\mu_R)}{\pi}\,\frac{\beta_0^{\mbox{\tiny SQCD}}}{2}\,\ln\left(\frac{\mu_R^2}{(m_{\tilde{g}}+\overline{m}_{\tilde{q}})^2}\right)\right]
\nonumber\\
&&\times\,\left(\delta(1-z)+\frac{\alpha_s(\mu_R)}{\pi}\,\mathcal{R}^{[R]}(z)\right)\,,
\label{eqn:FnloSG}
\end{eqnarray}
with
\begin{eqnarray}
\mathcal{R}^{[R]}(z)&=&(C_A+C_F)\left\{2\left[\frac{\ln(1-z)}{1-z}\right]_+-\ln\left(\frac{\mu_F^2}{(m_{\tilde{g}}+\overline{m}_{\tilde{q}})^2}\right)\left[\frac{1}{1-z}\right]_+\right\}
\nonumber\\
&&-C_R\left[\frac{1}{1-z}\right]_+-\delta(1-z)\,\frac{\beta_0^{\mbox{\tiny
      SQCD}}+3C_F}{2}\,\ln\left(\frac{\mu_F^2}{(m_{\tilde{g}}+\overline{m}_{\tilde{q}})^2}\right)
\,,
\label{eqn:FnloSG2}
\end{eqnarray}
where $\beta_0^{\mbox{\tiny SQCD}}=3C_A-2T_Fn_f$ is the one-loop
coefficient of the SUSY QCD beta function. The normalization factors
$\mathcal{N}_{ij}^{[R]}$ are given in Eq.~(\ref{eqn:NloSG}), the
constants $C_R$ in Tab.~\ref{tab:table1}. The conventional
plus-distribution\footnote{The plus-distribution follows the
prescription $\int_0^1dz\,\left[\frac{\ln^n(1-z)}{1-z}\right]_{+}f(z)
\equiv\int_0^1dz\,\frac{\ln^n(1-z)}{1-z}\left[f(z)-f(1)\right]$ for
$n=0,1,\ldots$ and any test function $f(z)$. If the lower integration
boundary is given by $\rho>0$ the plus distribution can be replaced by
the $\rho$-description via
$\left[\frac{\ln^n(1-z)}{1-z}\right]_{+}\rightarrow\frac{\ln^{n+1}(1-\rho)}{n+1}\delta(1-z)+\left[\frac{\ln^n(1-z)}{1-z}\right]_{\rho}$
where the latter is defined through
$\int_\rho^1dz\,\left[\frac{\ln^n(1-z)}{1-z}\right]_{\rho}f(z)
\equiv\int_\rho^1dz\,\frac{\ln^n(1-z)}{1-z} \left[f(z)-f(1)\right]$.}
has been employed to regularize singularities appearing at $z=1$. The
terms of the first line of Eq.~(\ref{eqn:FnloSG2}) stem from
initial-state radiation and are therefore proportinal to the sum of the
Casimirs of the adjoint and the fundamental representation. The first
term in the second line originates from final-state radiation and
depends on the representation of the gluino-squark state.  The last term
originates from the renormalization of the parton densities. We do not
include the estimate of the virtual corrections based on the colour
summed open production -- as performed in \cite{Hagiwara:2009hq} for the
gluino-gluino case.
\section{\label{sec:num}Numerical results}
\subsection{\label{subsec:SPS}Benchmark Points}
From now on we limit our discussion to the benchmark points (p) and
(q). First we focus on (p), where the gluino mass and the four squark
masses of the first generation lie in a range of less than
$30\,\mbox{GeV}$. For the right handed squarks the sum of the
decay widths is slightly smaller than the level spacing
$\Delta\overline{M}_{\tilde{g}\tilde{q}}$, for the left handed ones
approximately twice as big.  Thus, the enhancement from the
lowest lying resonance will be visible in the differential cross
section.
\par
As an alternative we consider benchmark point (q). In this case the
constituent decay rates are significantly larger than the level
spacing. Final state interaction will, nevertheless, lead to a
significant enhancement and distortion of the cross section in the
threshold region.
\subsection{\label{subsec:GreenNum}Green's Function}
As a characteristic scale for the strong coupling in the potential we
have chosen the Bohr radius
\begin{eqnarray}
\mu_G^{[R]}&\equiv&2\,m_{\rm red}\left|C^{[R]}\right|\,\alpha_s\bigl(\mu_G^{[R]}\bigr)
\,.
\label{eqn:muG}
\end{eqnarray}
\par
\begin{figure}[tb]
\begin{center}
\begin{tabular}{cc}
\includegraphics[angle=270,width=0.505\textwidth]{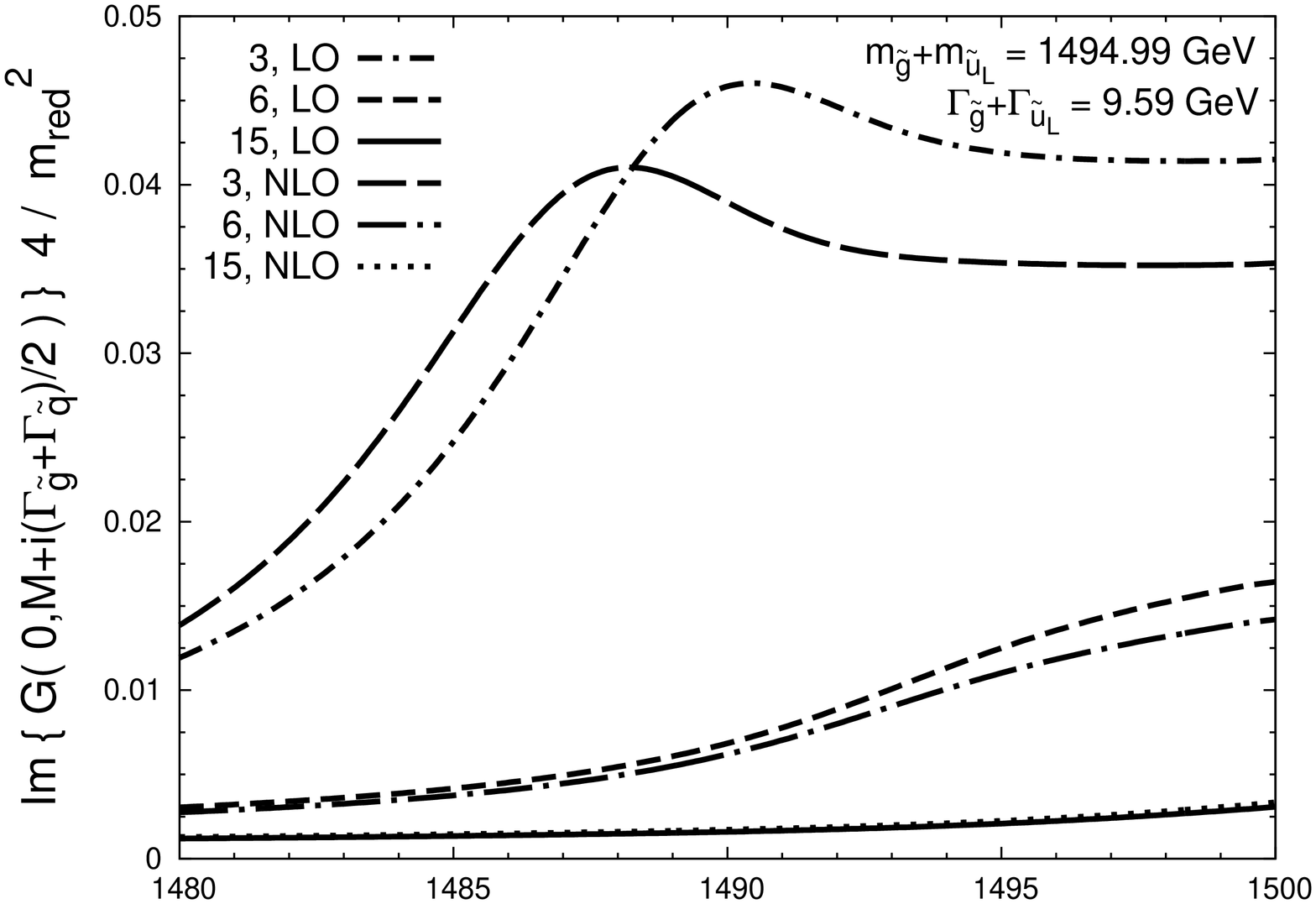}
&
\hspace{-.88cm}
\includegraphics[angle=270,width=0.505\textwidth]{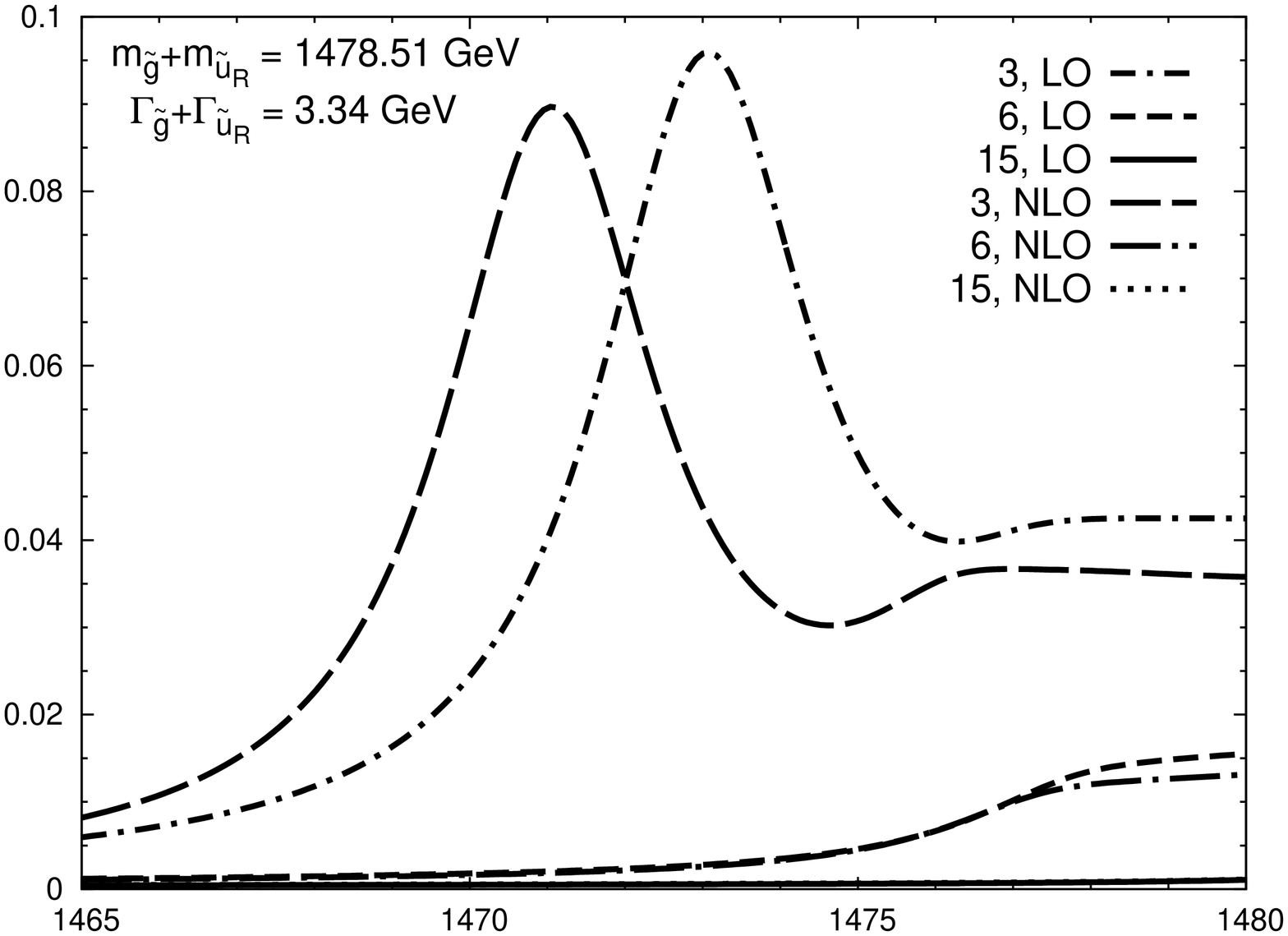}
\vspace{-.5cm}
\\
\includegraphics[angle=270,width=0.505\textwidth]{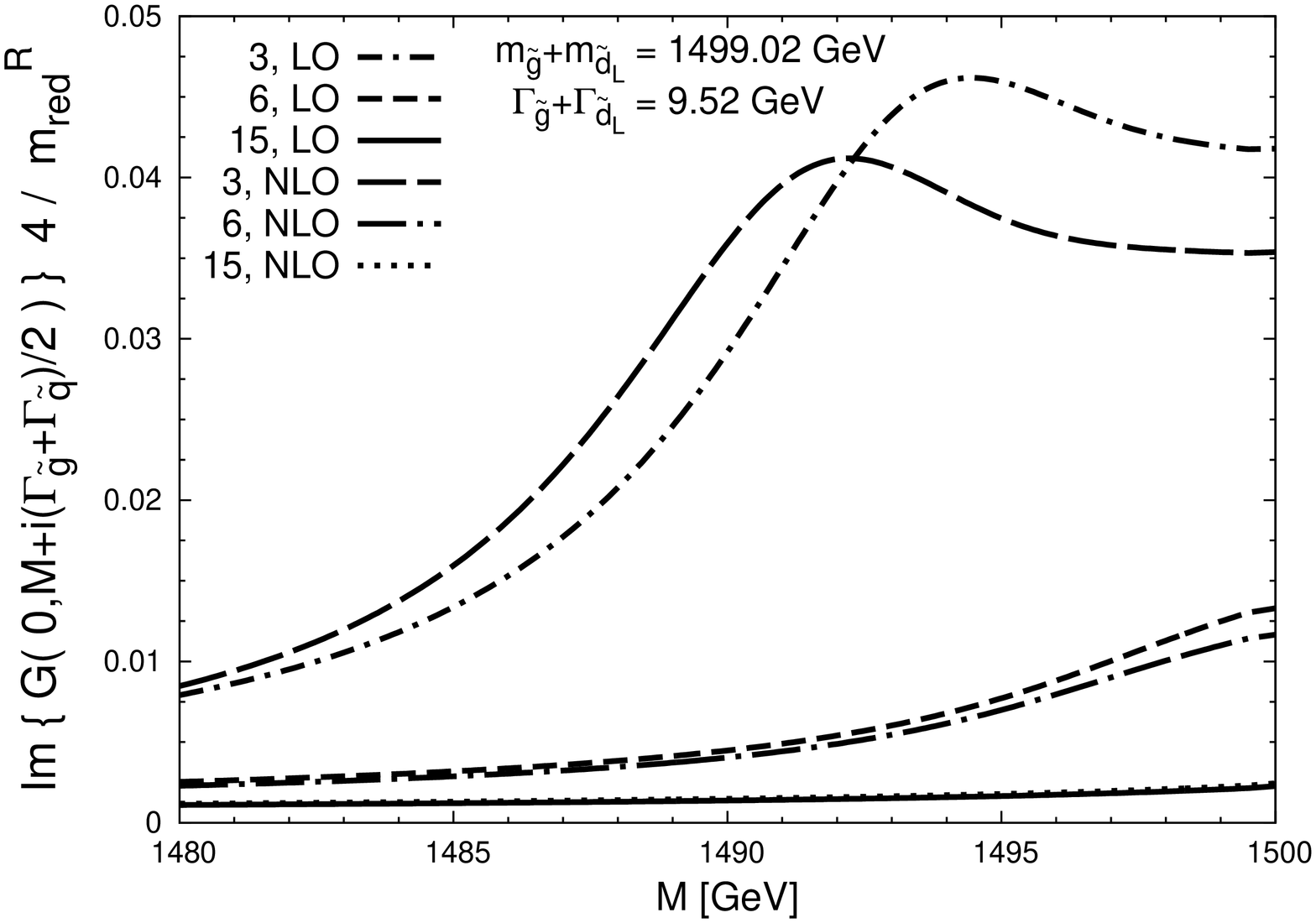}
&
\hspace{-1.cm}
\includegraphics[angle=270,width=0.505\textwidth]{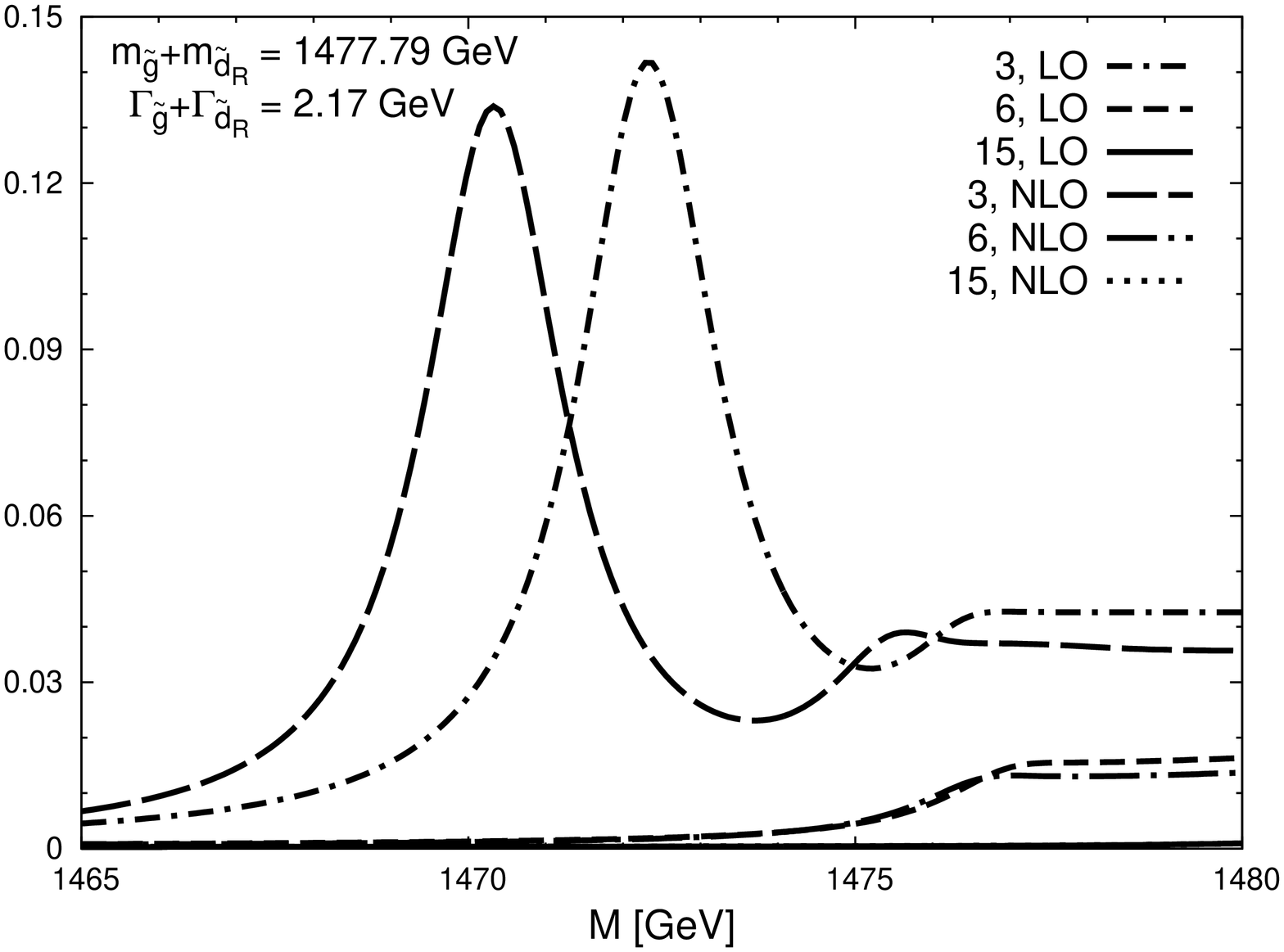}
\end{tabular}
\caption{Imaginary part of the Green's function for scenario (p) for the
  up-type (at the top) and the down-type (at the bottom) squarks. LO and
  NLO curves are plotted for all three colour configurations, whereas
  both curves lie on top of each other for the ${\bf 15}$ representation.}
\label{fig:Green}
\end{center}
\end{figure}
The normalized imaginary part of the Green's function at the origin is
displayed in Fig.~\ref{fig:Green}. The plots involving the thresholds
with up-type (down-type) squarks can be found at the top (at the bottom)
while left-handed (right-handed) squarks can be found on the left
(right) side. LO and NLO curves are displayed separately and the poor
convergence of the binding energies, which had been observed for 
quarks \cite{Penin:2002zv,Beneke:2005hg,Penin:2005eu} as well as for 
gluinos \cite{Kauth:2009ud}, results in a shift of $2\,\mbox{GeV}$
between the ground state of the most attractive colour-triplet
configuration going from LO to NLO. The effect of the larger decay
widths for the left-handed squarks manifests itself in the smearing of
this resonance compared to the right-handed particles. The attractive
potential of the colour-sextet representation is suppressed by a
factor of three such that no resonances can be seen. The contribution of
the repulsive ${\bf 15}$ configuration is negligible in the threshold region
and the LO and NLO curve lie on top of each other.
\subsection{\label{subsec:CrossNum}Cross section}
Combining the results of the previous section, the threshold behaviour
of gluino-squark production in proton-proton collision can be
computed. For the numerical evaluation of the strong coupling we used
the program {\tt RunDec} \cite{Chetyrkin:2000yt} and the starting values
$\alpha_s(M_Z)=0.13939$ for the MSTW2008LO PDFs and
$\alpha_s(M_Z)=0.12018$ for the MSTW2008NLO set
\cite{Martin:2009iq}. For $\mu_H$ (Eq.~(\ref{eqn:muH})) the running is
performed at the two-loop level. Decoupling of the top-quark at
$m_t=172.0\,\mbox{GeV}$ and decoupling of the gluino and squarks at
$\mu_H$ is performed at one-loop precision (see for example
\cite{Bauer:2008bj}).
\par
For simplicity the squark mass entering the reduced mass of the cross
section and the determination of the coupling is understood as averaged
mass of the squarks of the first generation from
Eq.~(\ref{eqn:Msqu1}). The information of the four different squarks is
only encoded in the decay widths $\Gamma_{\tilde{q}_i}$ and in the total
mass of the gluino-squark system entering the Green's function via the
relative velocity $v$ from Eq.~(\ref{eqn:Green3}) and determining the
position of the threshold.
\par
\begin{figure}[p]
\begin{center}
\begin{tabular}{cc}
\includegraphics[angle=270,width=0.505\textwidth]{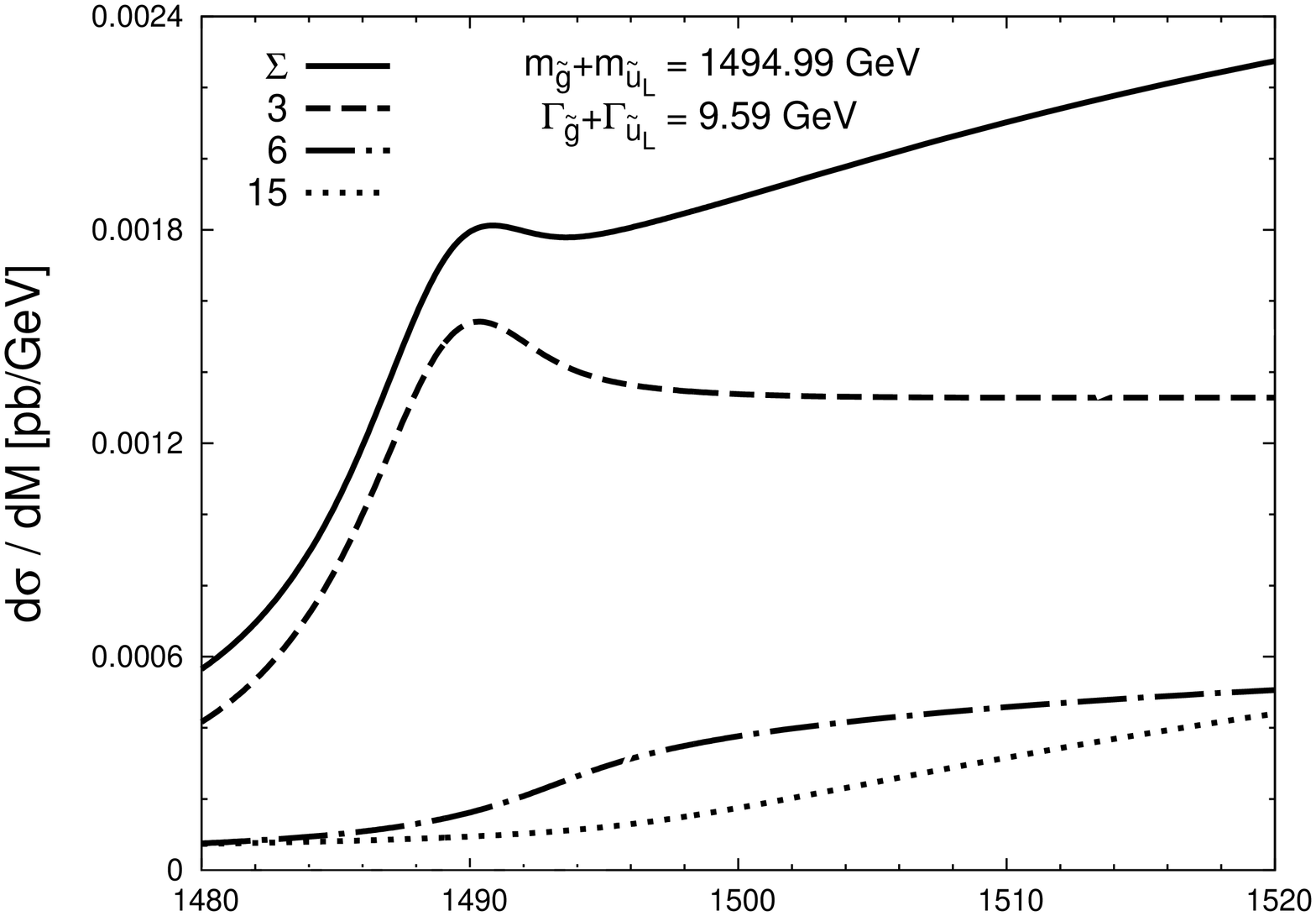}
&
\hspace{-.88cm}
\includegraphics[angle=270,width=0.505\textwidth]{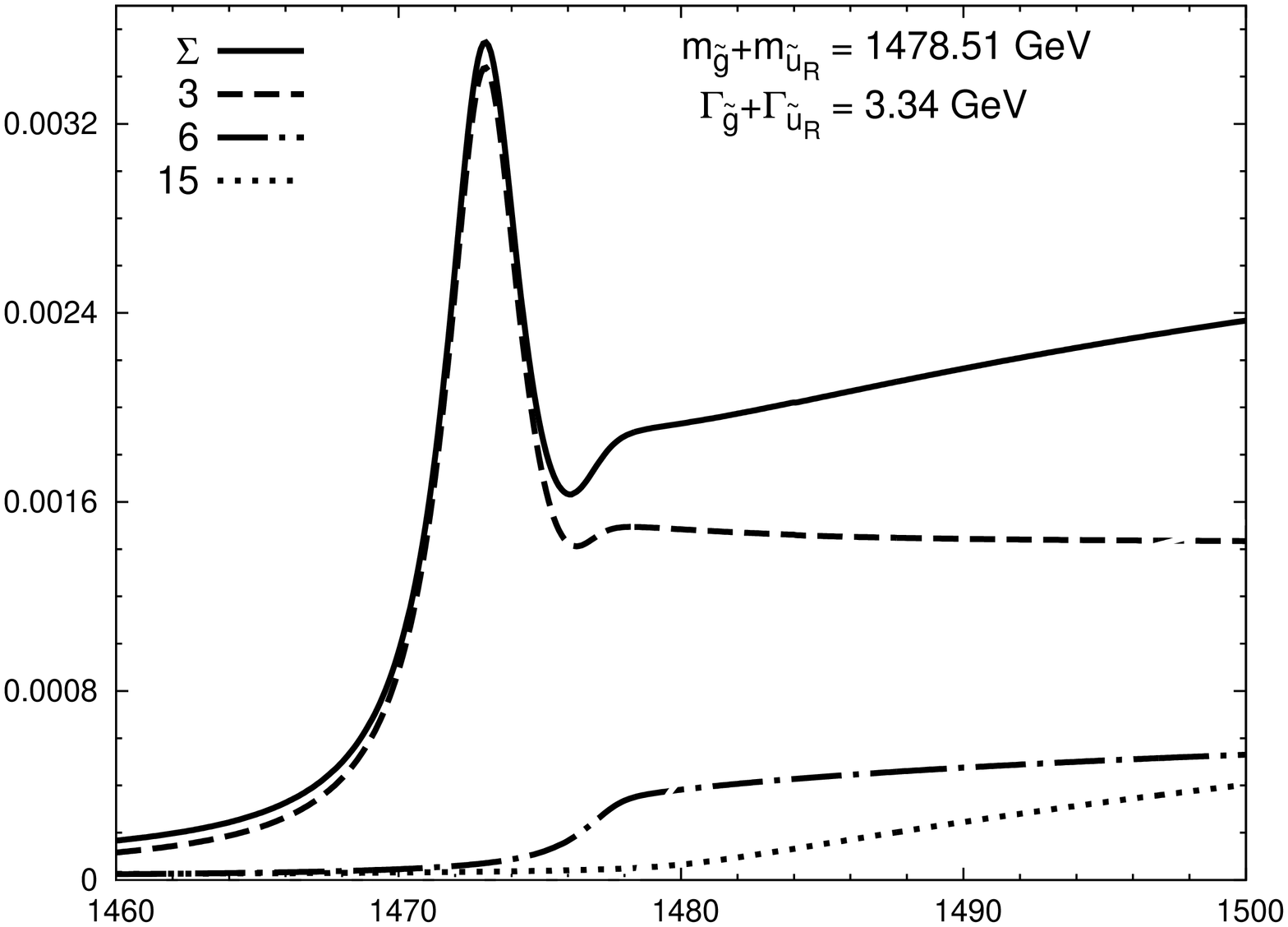}
\vspace{-.5cm}
\\
\includegraphics[angle=270,width=0.505\textwidth]{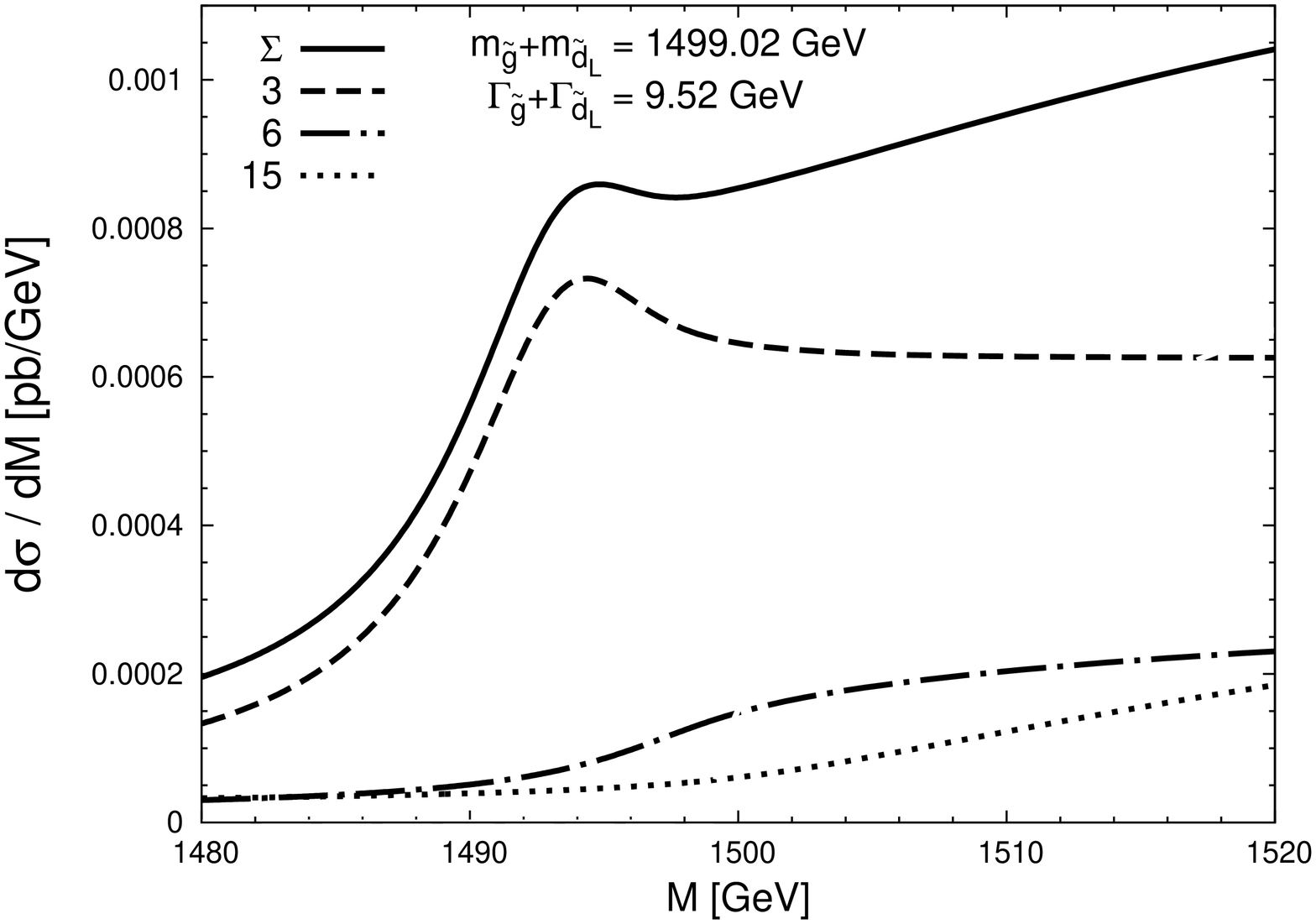}
&
\hspace{-1.cm}
\includegraphics[angle=270,width=0.505\textwidth]{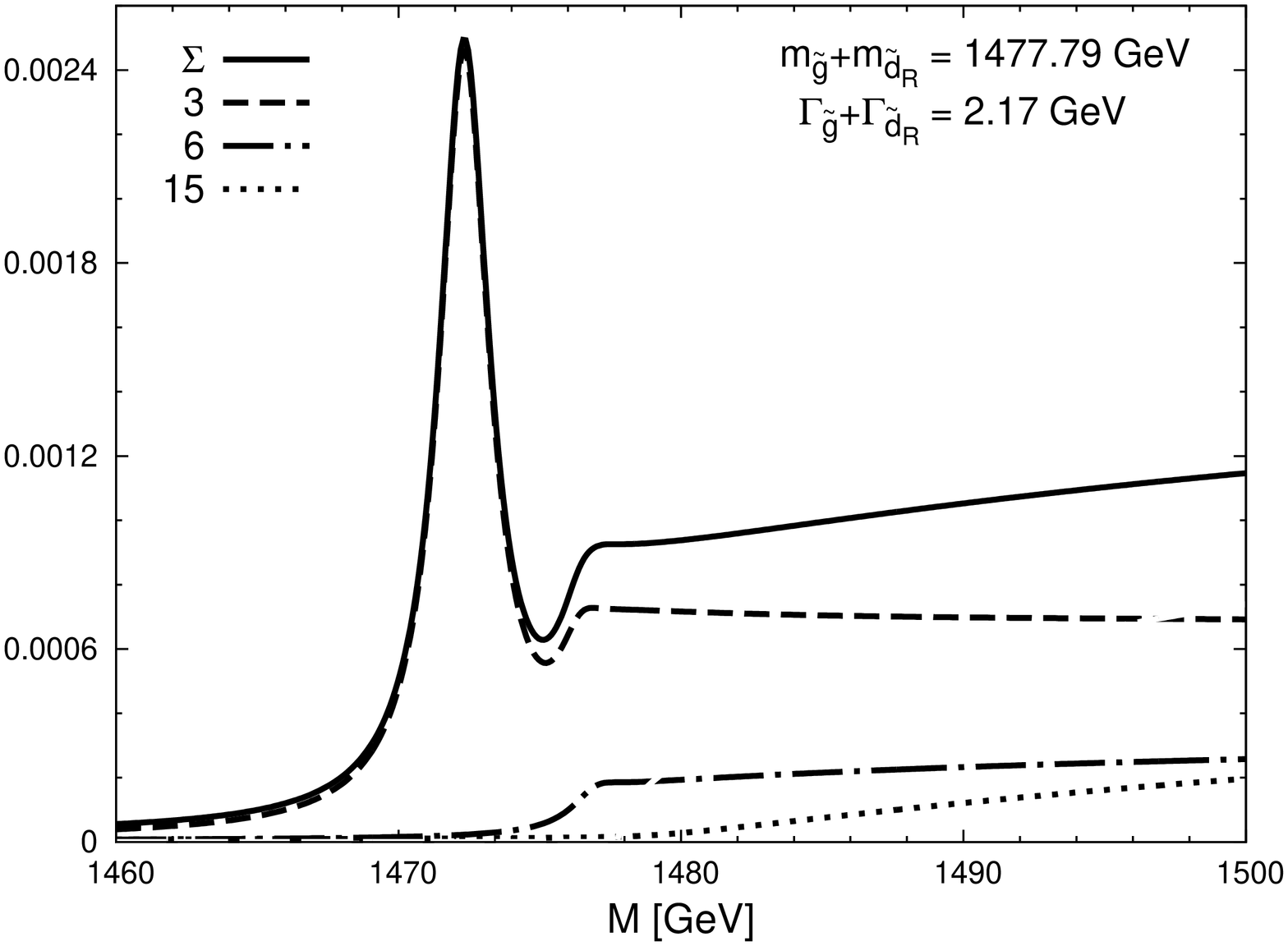}
\vspace{-0.3cm}
\\
&
\hspace{-9.cm}
\includegraphics[angle=270,width=.645\textwidth]{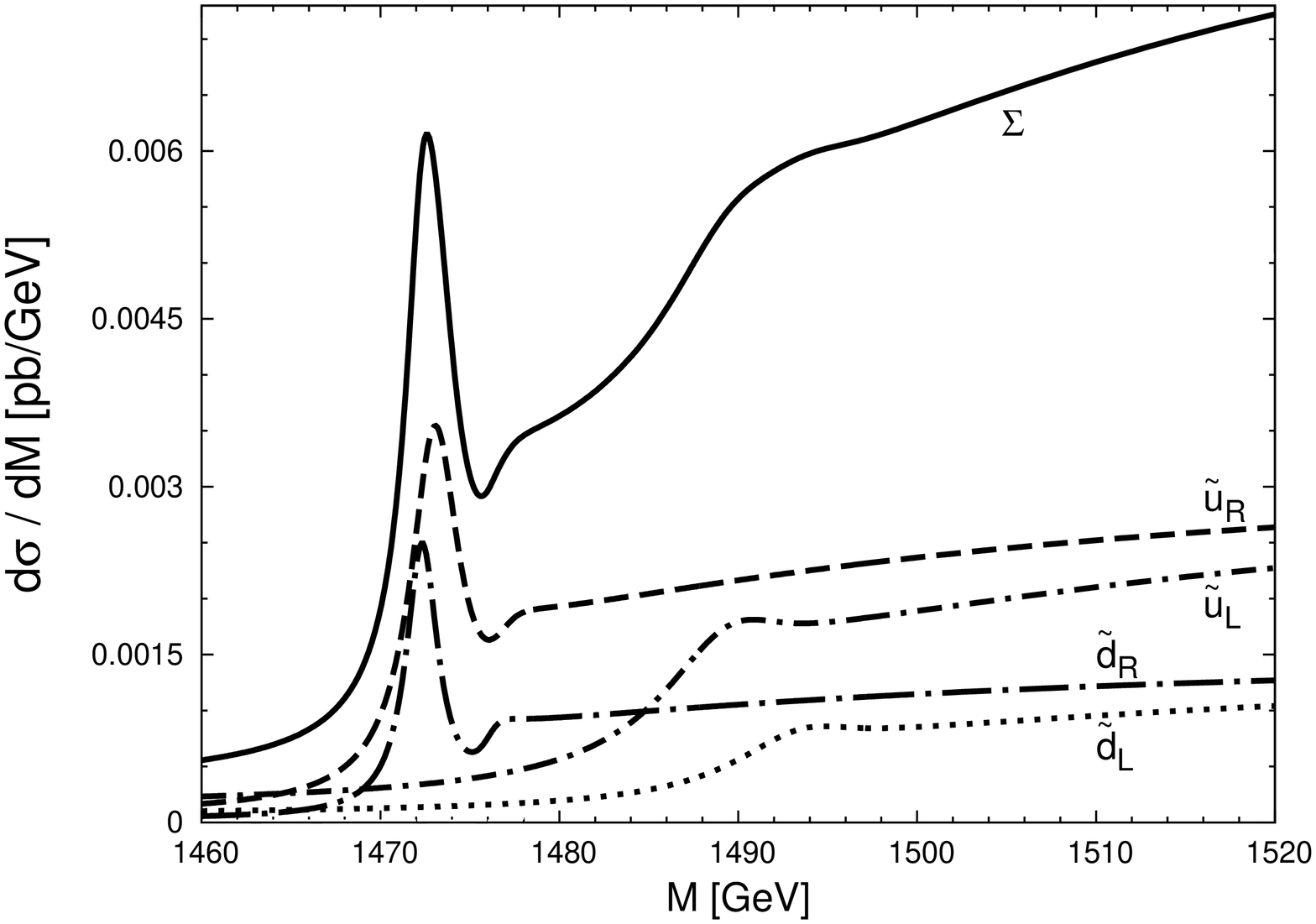}
\end{tabular}
\caption{Prediction for the invariant mass distribution for scenario (p)
  at LO. The upper four plots represent the results for the
  configurations separately and are ordered as in
  Fig.~\ref{fig:Green}. The figure at the bottom displays the individual
  contributions from $\tilde{u}_{L,R}$ and $\tilde{d}_{L,R}$ and their sum.}
\label{fig:CrossLO}
\end{center}
\end{figure}
In Fig.~\ref{fig:CrossLO} the LO result for the production cross
section, differential in the invariant mass of the two supersymmetric
particles, can be found. The contributions of the channels
$gq\rightarrow\tilde{g}\tilde{q}$ and
$g\overline{q}\rightarrow\tilde{g}\tilde{q}^*$ are both included. We
have employed the parton distribution functions (PDF) of MSTW2008LO
\cite{Martin:2009iq}. Both factorization and the renormalization scale
have been set to the hard scale
\begin{eqnarray}
\mu_F&=&\mu_R\hspace{0.2cm}=\hspace{0.2cm}\mu_H\hspace{0.2cm}\equiv\hspace{0.2cm}m_{\tilde{g}}+\overline{m}_{\tilde{q}}
\,.
\label{eqn:muH}
\end{eqnarray}
Up to NLO Green's function and hard kernel are separately
renormalization scale invariant, such that the cross section involves
the two scales $\mu_G^{[R]}$ and $\mu_H$.
\par
The upper four plots of Fig.~\ref{fig:CrossLO} are arranged as in
Fig.~\ref{fig:Green}. They show the contributions from the three
representations and their sum. In the lower plot the contributions of
the four gluino-squark channels are shown as well as their sum. It is
clearly visible that, due to MFV and the quark content of the proton,
the contributions from up-type squarks are twice as big as the ones of
the curves involving down-type squarks. The resonances of the
right-handed squarks nearly coincide so that a relatively sharp
resonance will be visible in the invariant mass distribution.
\par
For degenerated squark and gluino masses, as it is roughly realized in
this scenario (p), the normalization factors of Eq.~(\ref{eqn:NloSG})
are comparable for all three colour representations. Therefore, the
relative magnitudes of the cross sections are largely governed by the
Green's function. Furthermore, the hard kernel does not vary strongly in
in the threshold region, hence the shapes of the curves shown in
Figs.~\ref{fig:Green} and \ref{fig:CrossLO} exhibit a similar behaviour.
\par
Let us now consider the (approximate) NLO result.  The NLO values for
the Green's function will be used together with the approximation of
Eqs.~(\ref{eqn:FnloSG}) and (\ref{eqn:FnloSG2}) for the hard part of the
invariant mass distribution.
\par
\begin{figure}[p]
\begin{center}
\begin{tabular}{c}
\includegraphics[angle=270,width=0.85\textwidth]{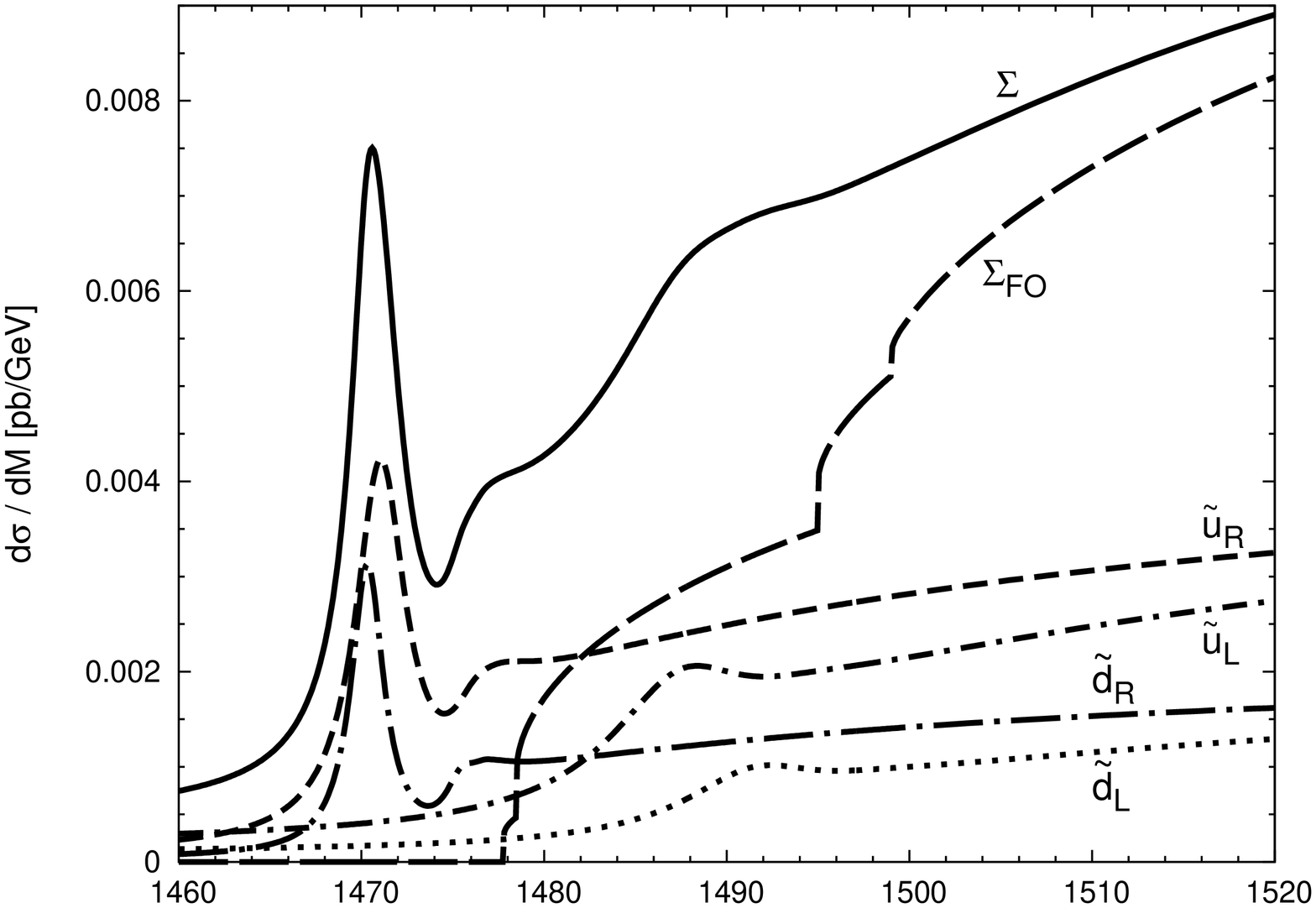}
\vspace{-0.8cm}
\\
\hspace{-.25cm}
\includegraphics[angle=270,width=0.85\textwidth]{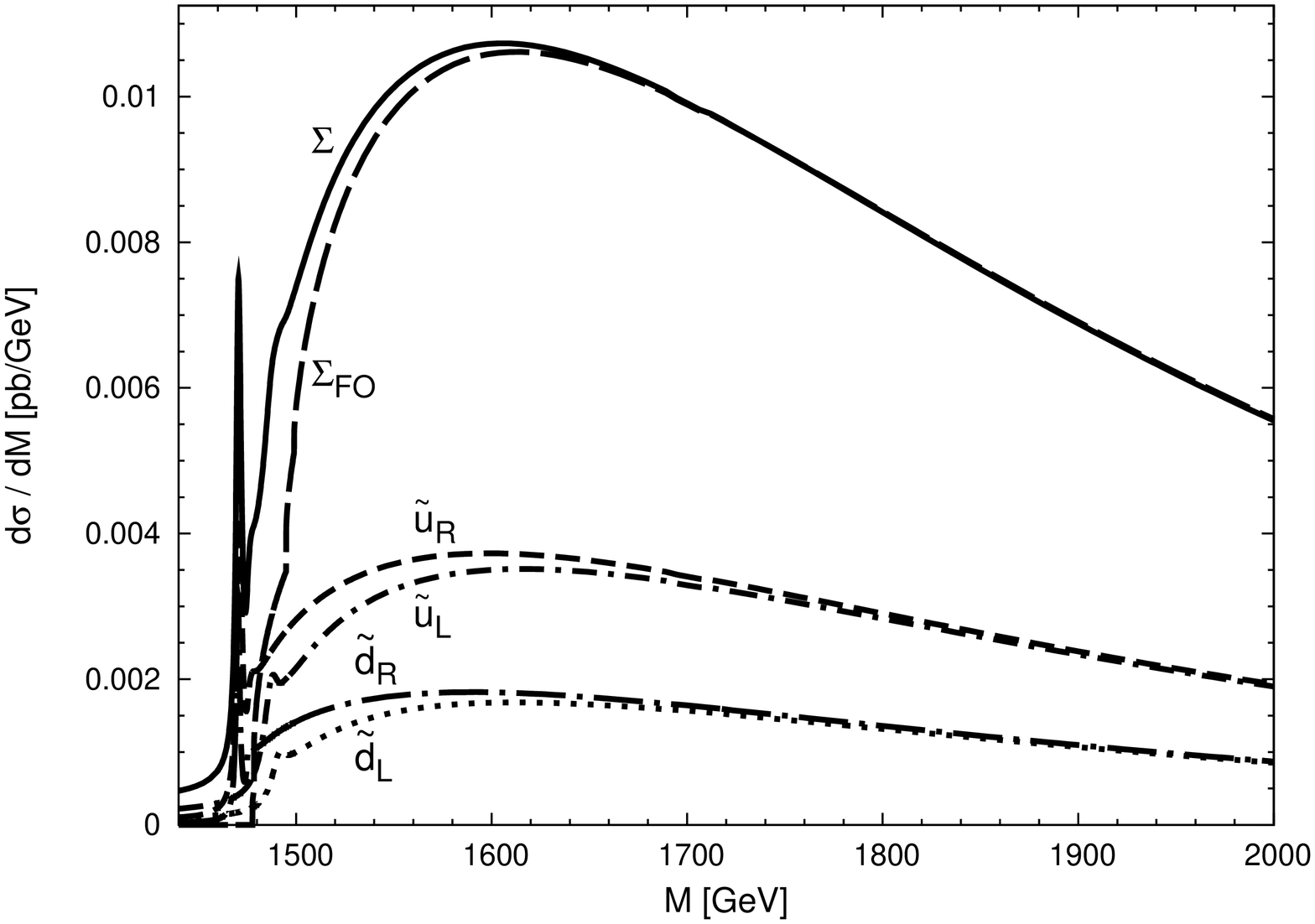}
\end{tabular}
\caption{Prediction for differential cross section as function of the
  invariant mass for scenario (p) evaluated in NLO approximation as
  defined in Eqs.~(\ref{eqn:FnloSG}) and (\ref{eqn:FnloSG2}) for the
  four squark species. Also shown is the sum of the four squarks (solid
  curve) and the FO prediction (long dashed curve).}
\label{fig:crossNLOsg}
\end{center}
\end{figure}
The result for benchmark point (p) is displayed in
Fig.~\ref{fig:crossNLOsg} for two regions of the invariant mass. The
shapes of LO (Fig.~\ref{fig:CrossLO}) and NLO predictions
(Fig.~\ref{fig:crossNLOsg}) look quite similar, differing mainly in
their relative normalization by about $20\%$. Furthermore, the
difference between the masses of $\tilde{u}_L$ and $\tilde{u}_R$,
respectively, becomes irrelevant for high energies, hence the difference
between the configurations with left- and righthanded squarks vanishes
far above the threshold. The contribution of the
$\tilde{g}\tilde{u}_{L/R}$ final states is twice the contribution of the
down-type squarks due to the structure of the proton. We also show the
sum of the fixed order (FO) curves where the Green's function is
replaced by its leading perturbative expansion in $\alpha_s$,
furthermore, in the limit
$\Gamma_{\tilde{g}},\Gamma_{\tilde{q}_i}\rightarrow 0$. Hence one gets
\begin{eqnarray}
\mbox{Im}\,\left\{G^{[R]}\left(M+\frac{i}{2}\left(\Gamma_{\tilde{g}}+\Gamma_{\tilde{q}_i}\right)\right)\right\}&\longrightarrow&\frac{m_{\rm
  red}^2}{\pi}\,v\left(1+C^{[R]}\,\frac{\alpha_s\,\pi}{2\,v}\right)\,.
\label{eqn:GreenFOsg}
\end{eqnarray}
with $v$ as in Eq.~(\ref{eqn:Green3}). In the threshold region the four
different values of the sum $m_{\tilde{g}}+m_{\tilde{q}_i}$ are clearly
visible from the kinks of the dashed FO curve. Above about
$M=1700\,\mbox{GeV}$ the two curves of the FO ansatz and the one using
the full Green's function roughly coincide and binding effects are
no longer of importance. The integrated
difference between the two functions
\begin{eqnarray}
\Delta\sigma&\equiv&\int_{M_0}^{M_1}\mbox{d}M\left[\left(\frac{\mbox{d}\sigma}{\mbox{d}M}\right)_{\mbox{\scriptsize
      Green}}-\left(\frac{\mbox{d}\sigma}{\mbox{d}M}\right)_{\mbox{\scriptsize
      FO}}\,\right]
\,,
\label{eqn:diff}
\end{eqnarray}
for $M_0=1415\,\mbox{GeV}$ and $M_1=1650\,\mbox{GeV}$ represents a
measure for the effect of the final state interaction and amounts to a
relative enhancement of about $2.57\%$ compared to the total cross
section of $\sigma_{\rm tot}=7.29\,\mbox{pb}$ which has been calculated
at NLO using {\tt Prospino2} \cite{Beenakker:1996ch}. It should be
mentioned that also effects of squarks from the second and third
generation are included in the total cross section.
\par
\begin{figure}[p]
\begin{center}
\begin{tabular}{c}
\includegraphics[angle=270,width=0.87\textwidth]{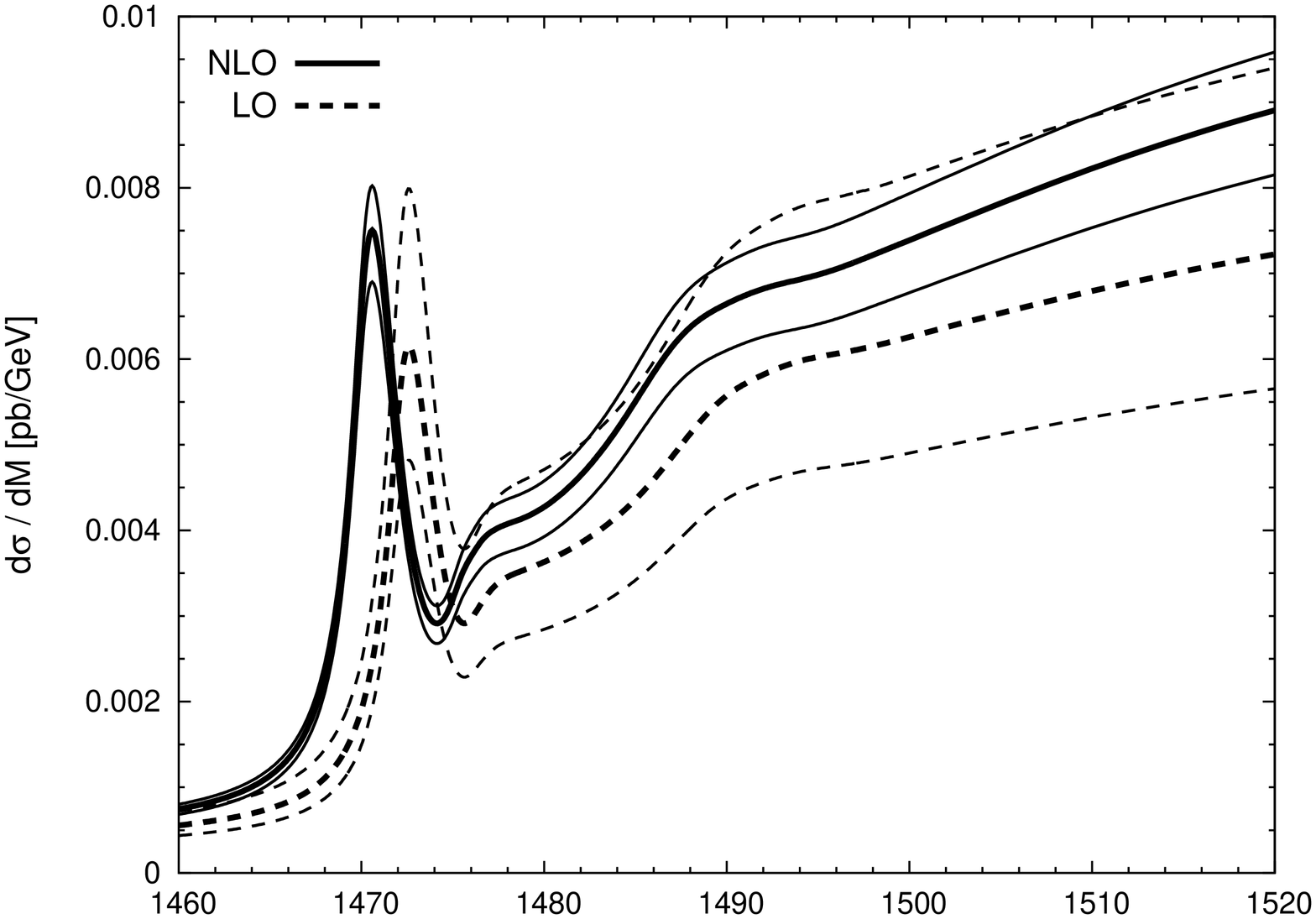}
\vspace{-0.8cm}
\\
\hspace{-.25cm}
\includegraphics[angle=270,width=0.87\textwidth]{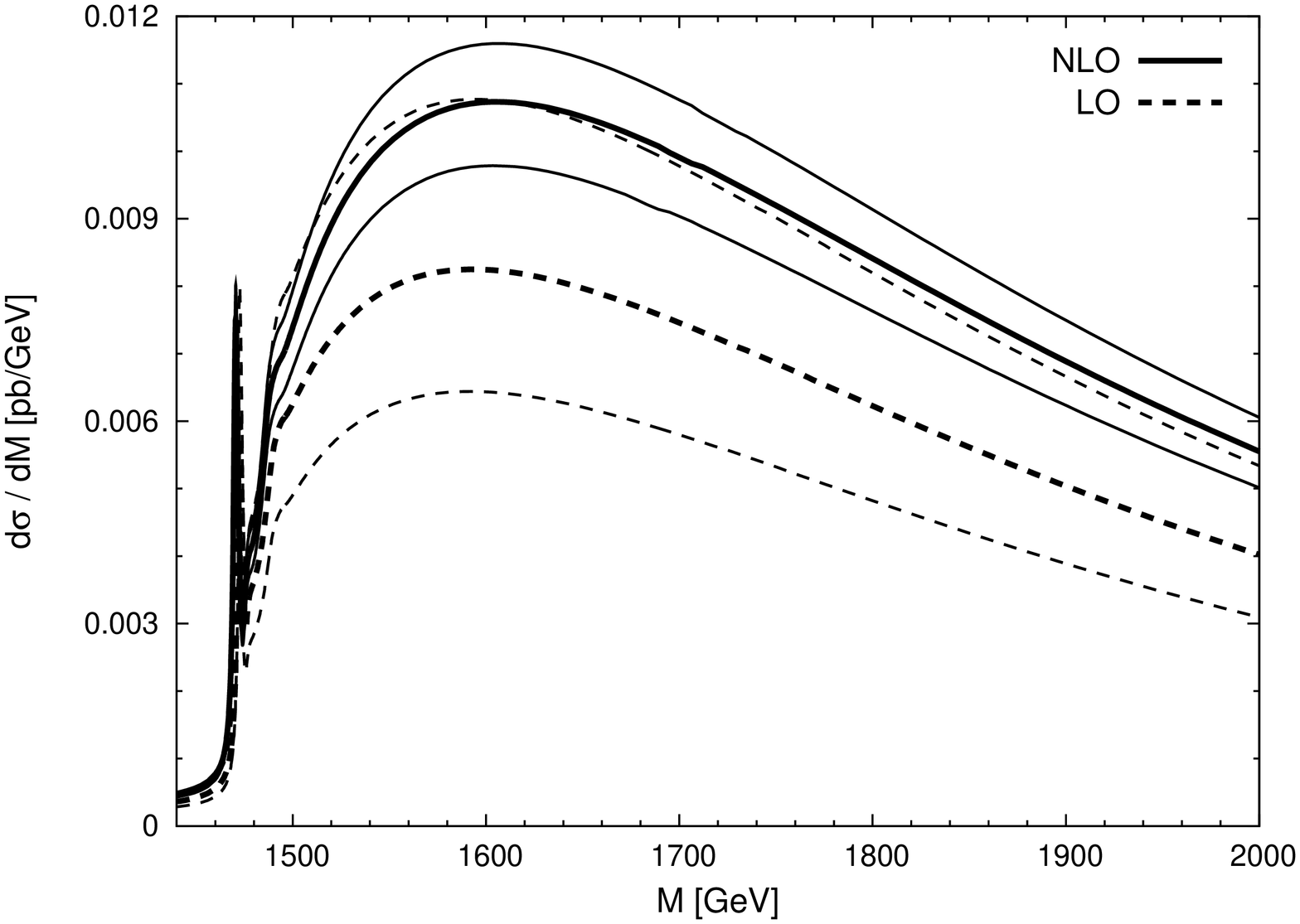}
\end{tabular}
\caption{Dependence of the invariant mass distribution on the choice of
  factorization and renormalization scale for scenario (p). For the
  scale of the Green's function Eq.~(\ref{eqn:muG}) has been adopted
  throughout.}
\label{fig:crossNLOmuSG}
\end{center}
\end{figure}
The sensitivity of our results on the factorization scale $\mu_F$ and
the renormalization scale $\mu_R$ is shown in
Fig.~\ref{fig:crossNLOmuSG}, both for LO and NLO predictions. The scales
$\mu_F$ and $\mu_R$ are chosen to be equal and varied between $\mu_H/2$
and $2\mu_H$ (see Eq.~\ref{eqn:muH}). As expected, the scale dependence
is significantly reduced for the NLO approximation, amounting to less
than $\pm10\%$. An estimate of the theoretical uncertainty may either be
based on this scale dependence, or, more conservatively, on the
difference between LO and NLO results, amounting to roughly $20\%$.
\par
\begin{figure}[tb]
\begin{center}
\begin{tabular}{cc}
\includegraphics[angle=270,width=0.505\textwidth]{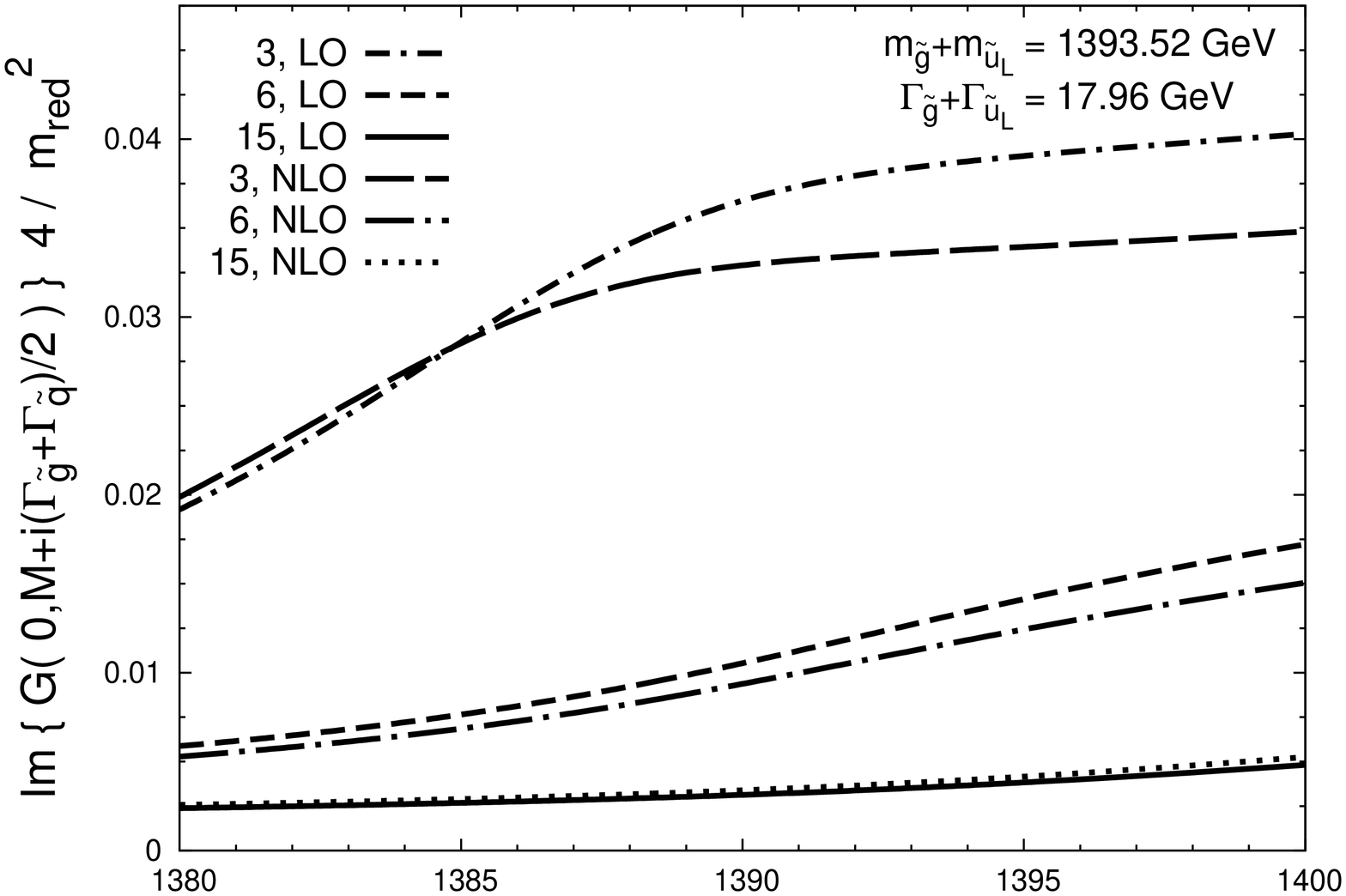}
&
\hspace{-.88cm}
\includegraphics[angle=270,width=0.505\textwidth]{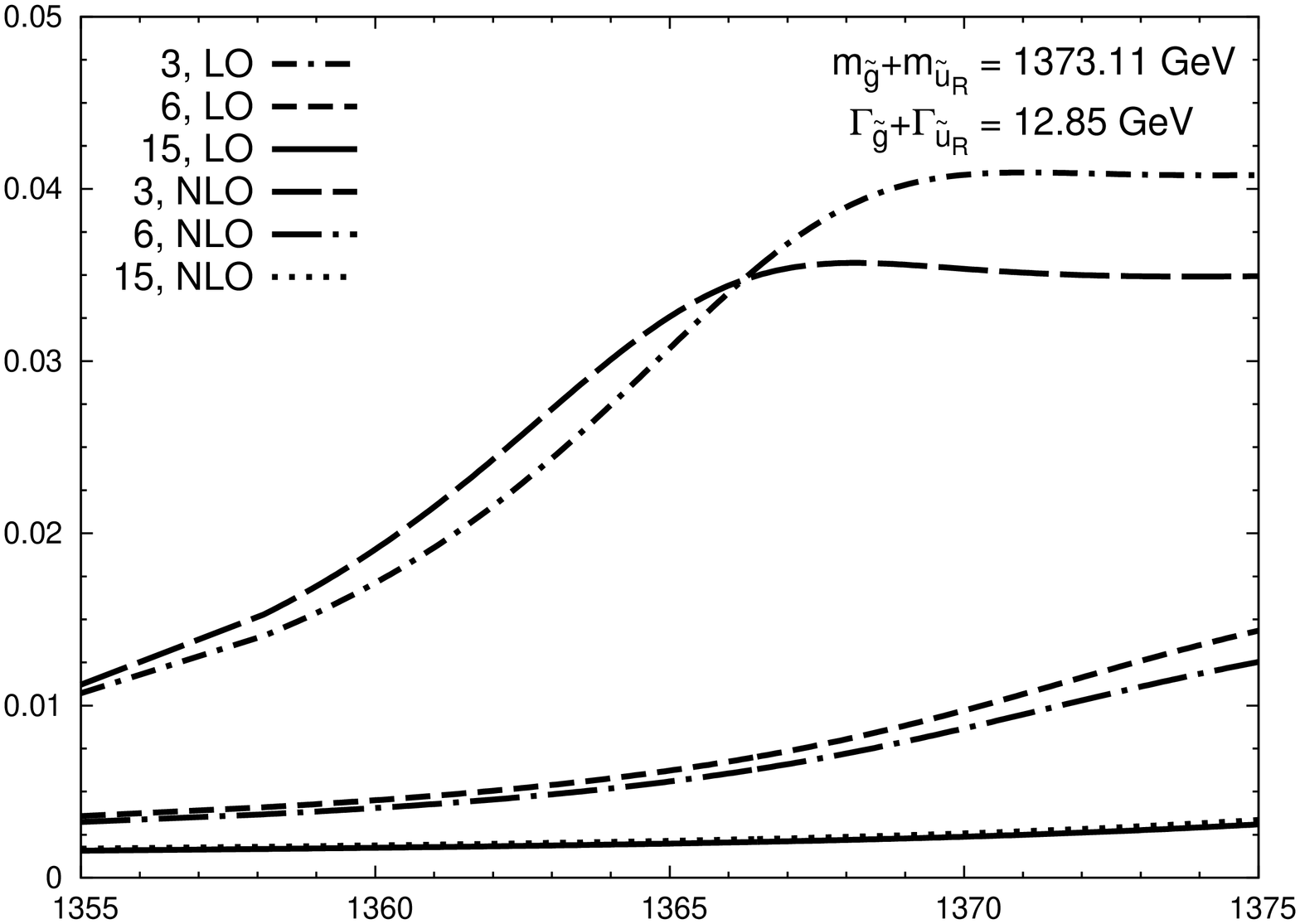}
\vspace{-.5cm}
\\
\includegraphics[angle=270,width=0.505\textwidth]{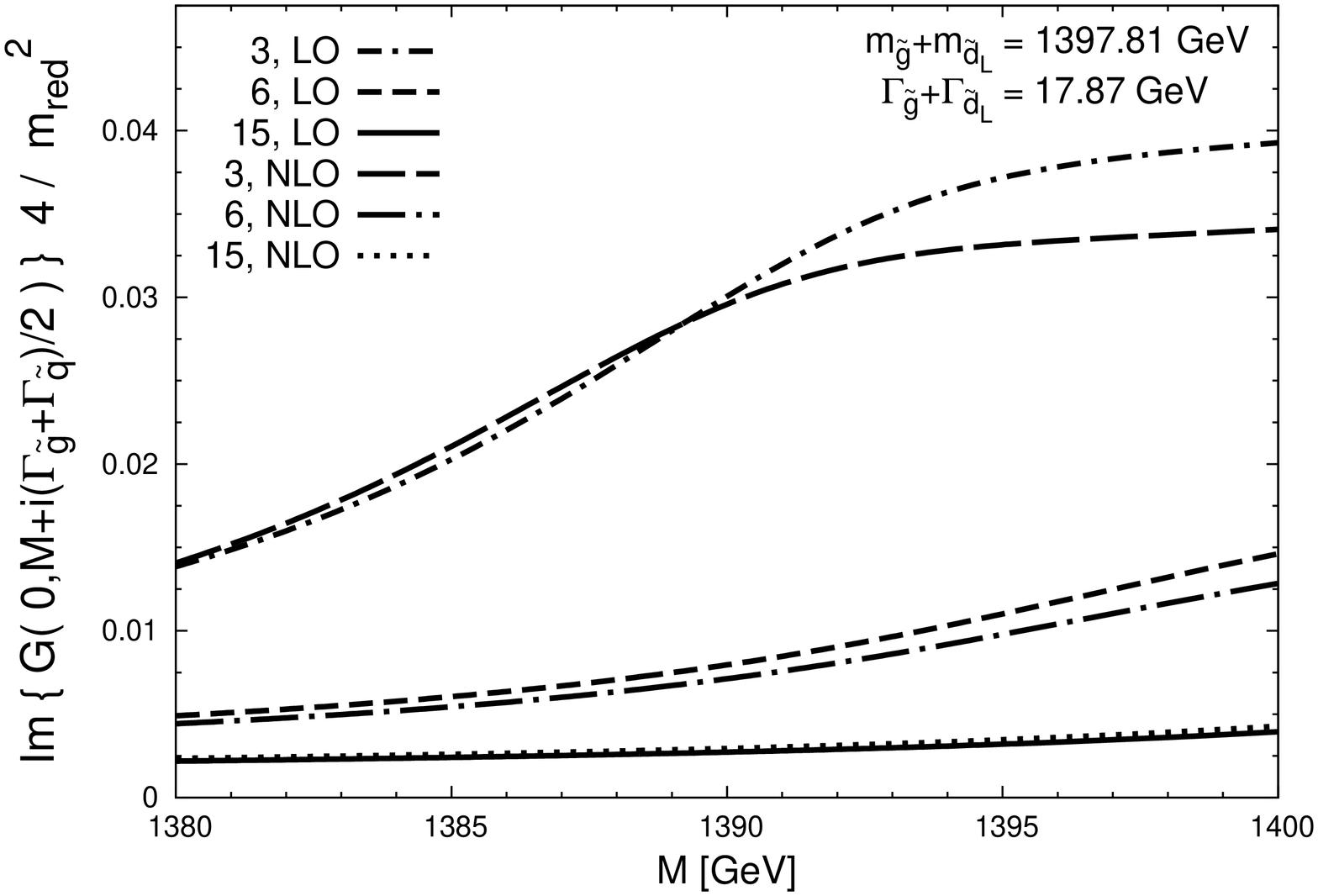}
&
\hspace{-1.cm}
\includegraphics[angle=270,width=0.505\textwidth]{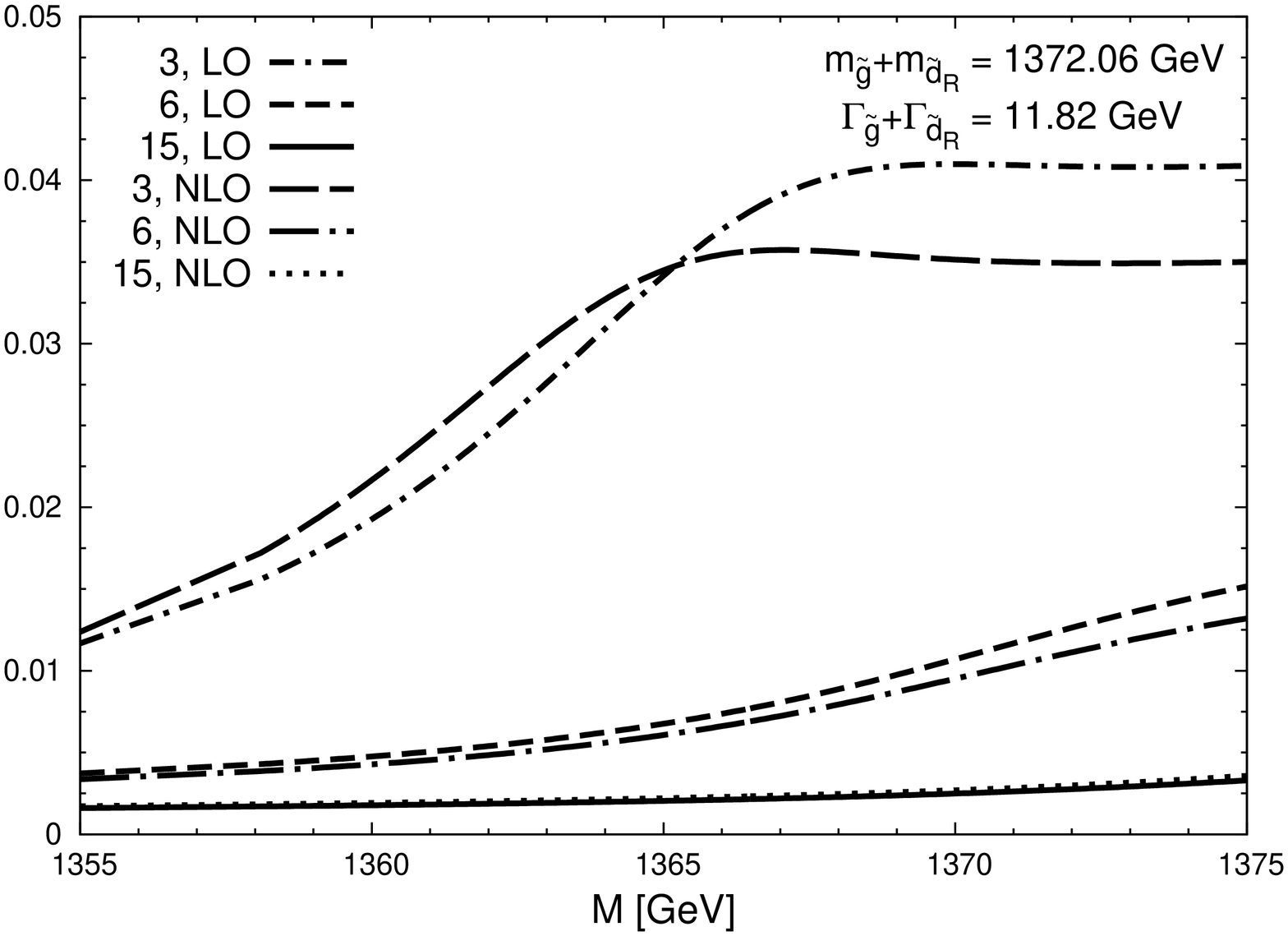}
\end{tabular}
\caption{Imaginary part of the Green's function for scenario (q) for the
  up-type (at the top) and the down-type (at the bottom) squarks. LO and
  NLO curves are plotted for all three colour configurations, whereas
  both curves lie on top of each other for the ${\bf 15}$ representation.}
\label{fig:QGreen}
\end{center}
\end{figure}
As a second option we study the prediction for benchmark point (q) with
constituent decay rates significantly larger than the level spacing. The
Green's functions are shown in Fig.~\ref{fig:QGreen}. The strong
enhancement of the triplet Green's function is evident, as well as the
strong suppression for the ${\bf 15}$ representation.
\begin{figure}[p]
\begin{center}
\begin{tabular}{cc}
\includegraphics[angle=270,width=0.505\textwidth]{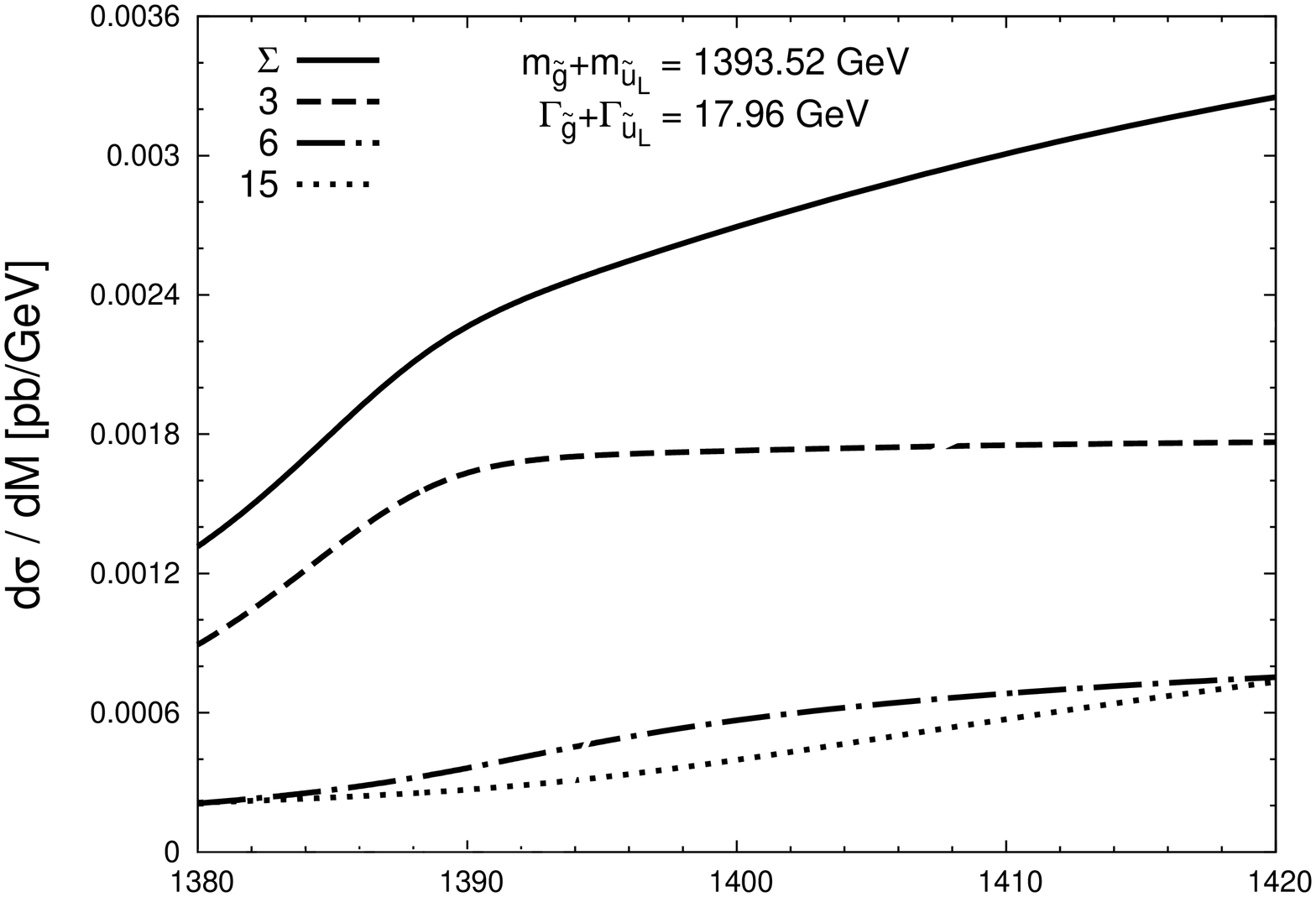}
&
\hspace{-.82cm}
\includegraphics[angle=270,width=0.505\textwidth]{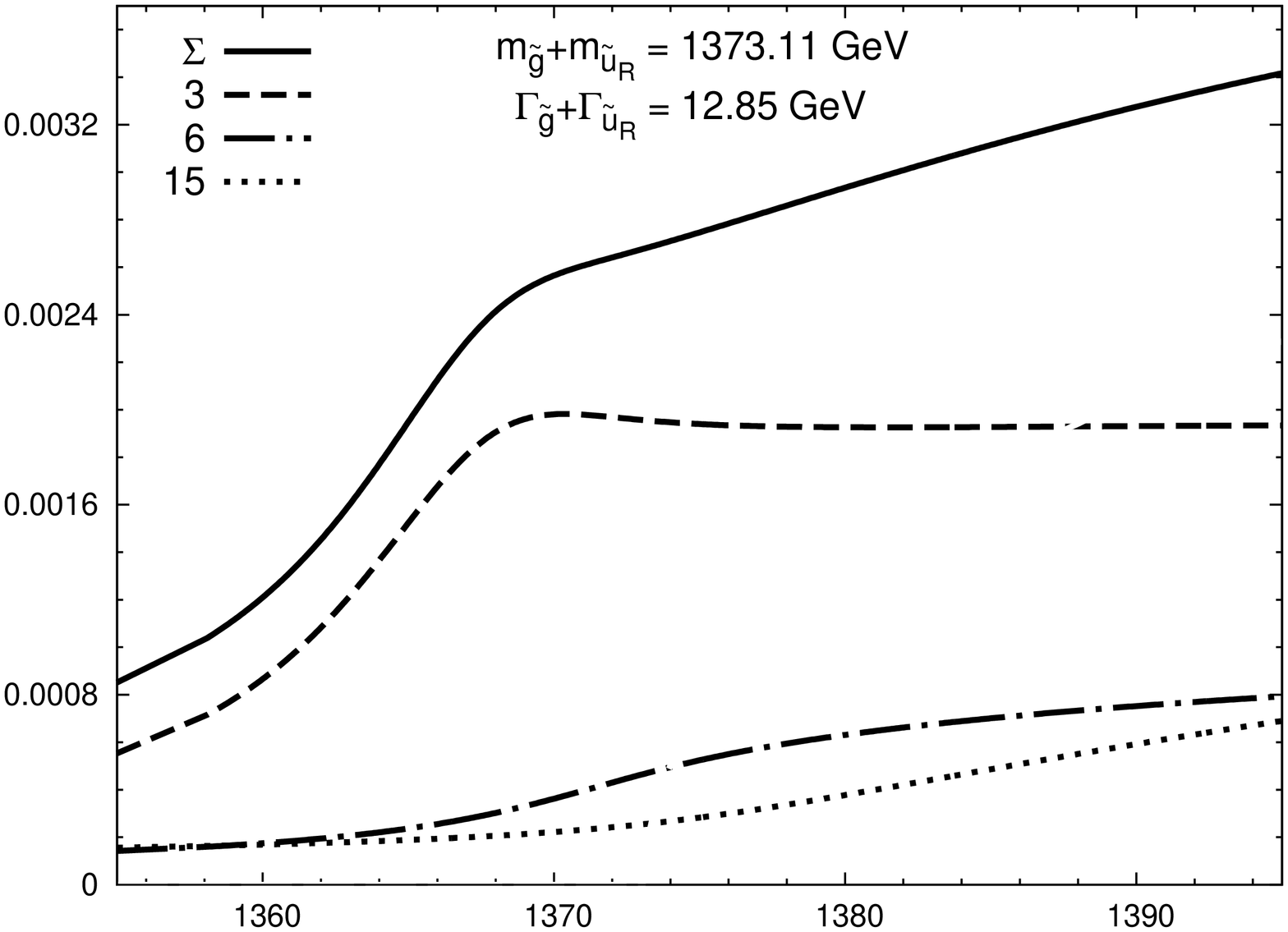}
\vspace{-.5cm}
\\
\includegraphics[angle=270,width=0.505\textwidth]{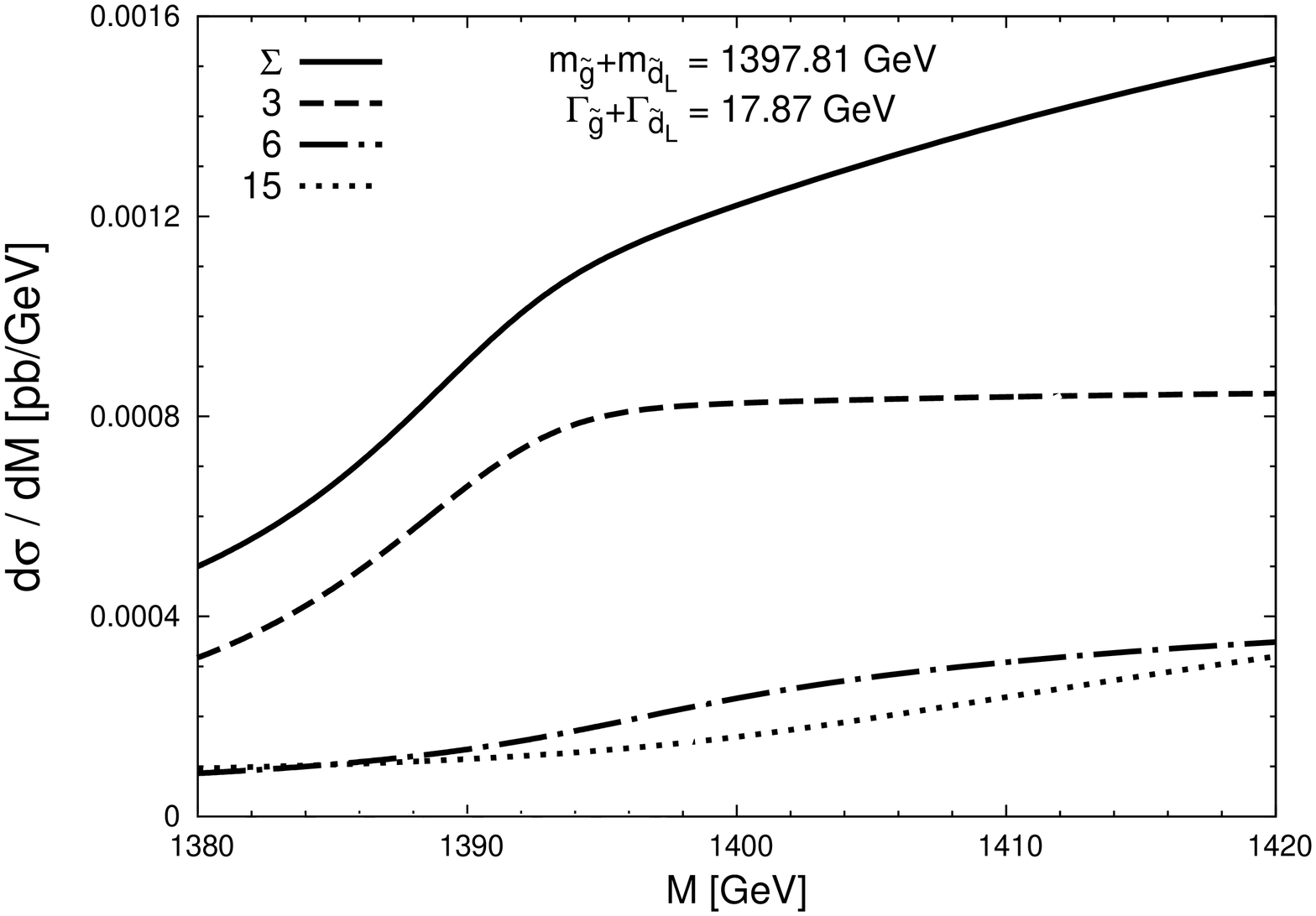}
&
\hspace{-0.84cm}
\includegraphics[angle=270,width=0.505\textwidth]{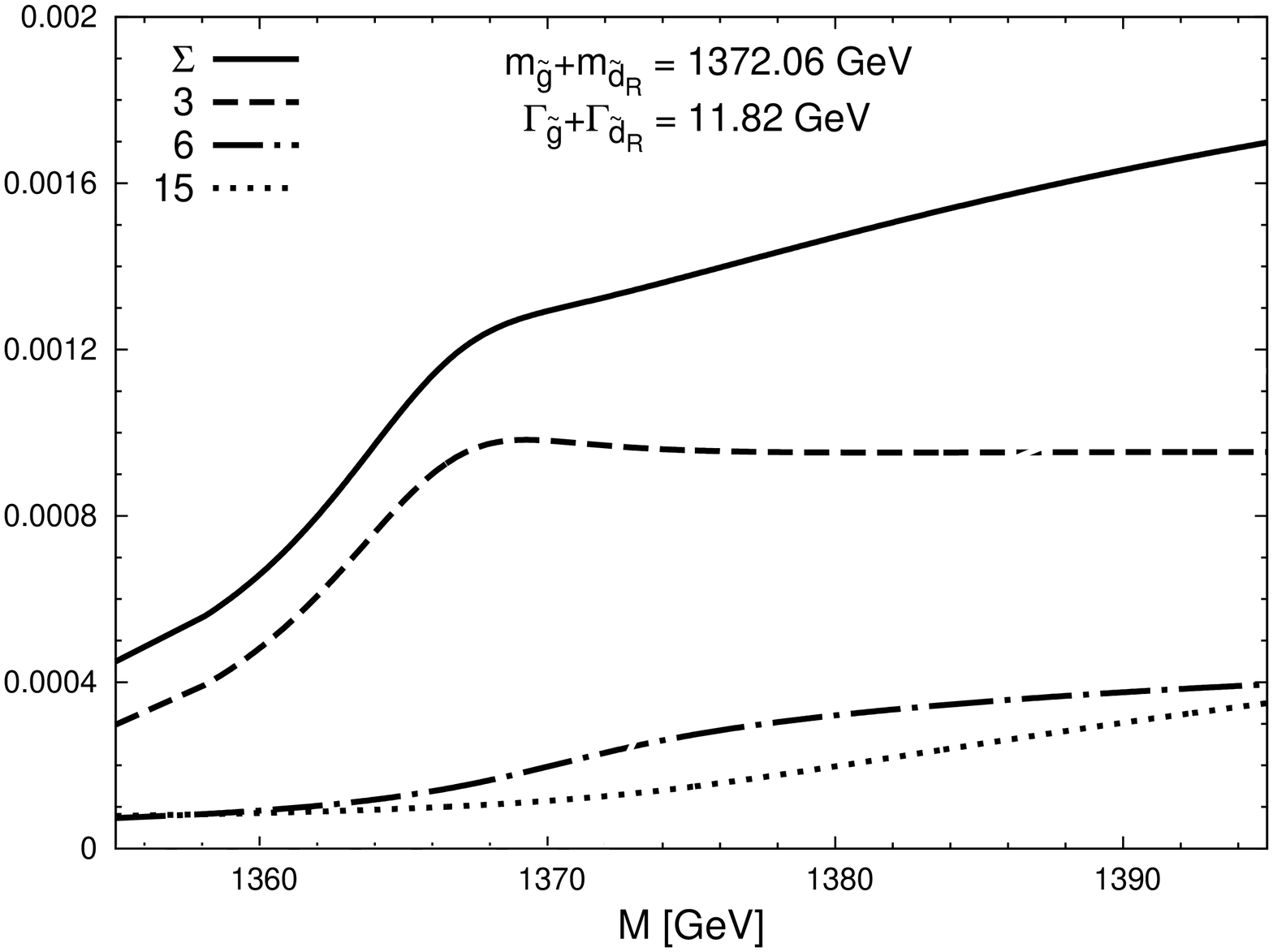}
\vspace{-0.3cm}
\\
&
\hspace{-9.cm}
\includegraphics[angle=270,width=.625\textwidth]{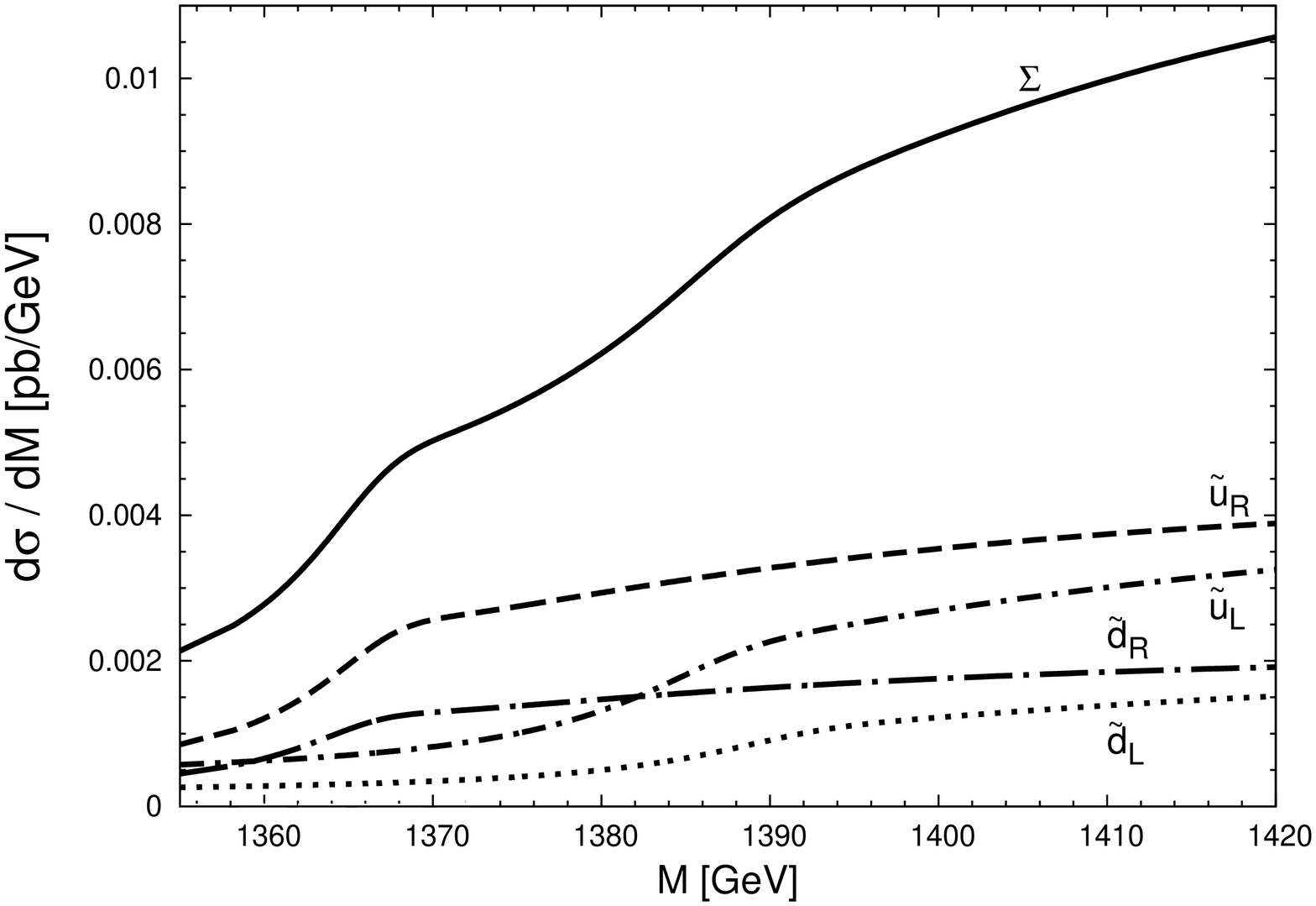}
\end{tabular}
\caption{Prediction for the invariant mass distribution for scenario (q)
  at LO. The upper four plots represent the results for the
  configurations separately and are ordered as in
  Fig.~\ref{fig:Green}. The figure at the bottom displays the individual
  contributions from $\tilde{u}_{L,R}$ and $\tilde{d}_{L,R}$ and their
  sum. Furthermore, the formalism of the Green's function is compared to
  the FO method.}
\label{fig:QCrossLO}
\end{center}
\end{figure}
The predictions for the differential cross sections are shown in
Fig.~\ref{fig:QCrossLO}, again separately for the different squark
species and the different representation. The step in
$\mbox{d}\sigma/\mbox{d}M$ slightly below the nominal threshold is still
visible for the individual squark species, it is, however, smeared out
when one considers the sum of all four channels. Nevertheless, again one
observes a significant enhancement of the differential cross section in
the threshold region, which extends from approximately $20\,\mbox{GeV}$
below to $50\,\mbox{GeV}$ above the nominal threshold. 
\par
\begin{figure}[p]
\begin{center}
\begin{tabular}{c}
\includegraphics[angle=270,width=0.87\textwidth]{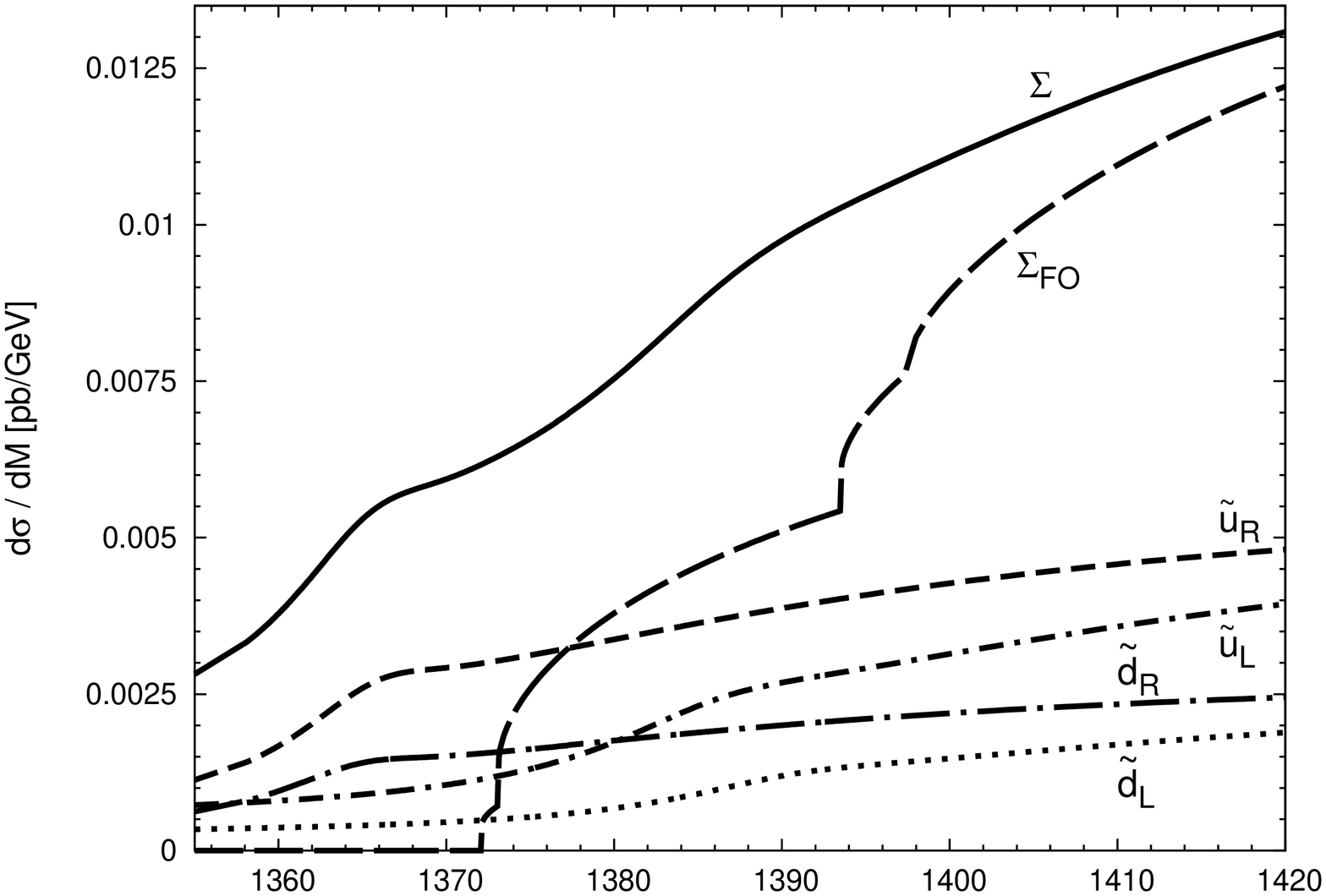}
\vspace{-0.8cm}
\\
\hspace{-.25cm}
\includegraphics[angle=270,width=0.87\textwidth]{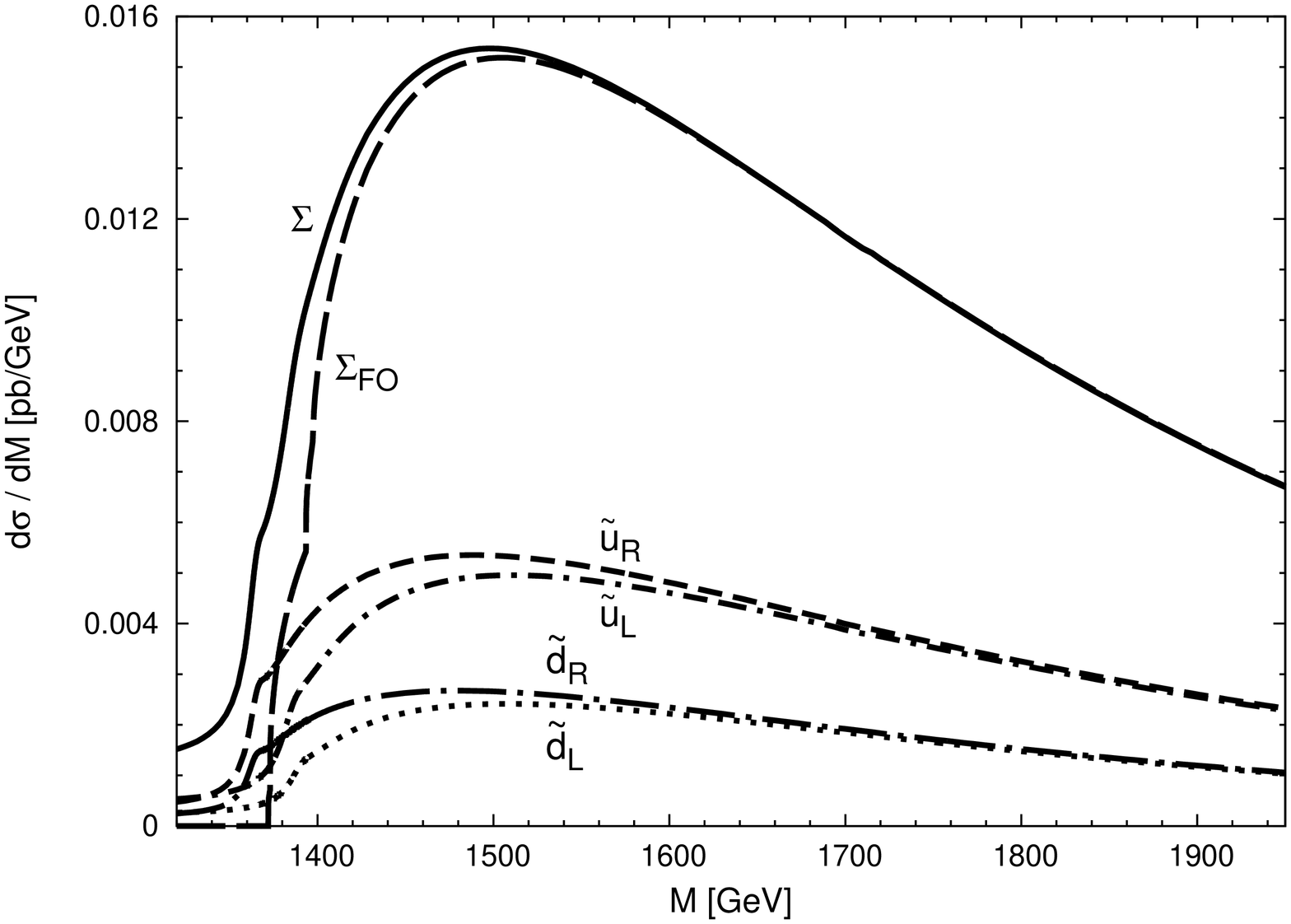}
\end{tabular}
\caption{Prediction for differential cross section as function of the
  invariant mass for scenario (q). The NLO approximation of
  Eqs.~(\ref{eqn:FnloSG}) and (\ref{eqn:FnloSG2}) has been adopted.}
\label{fig:QcrossNLOsg}
\end{center}
\end{figure}
The NLO result for benchmark point (q) is displayed in
Fig.~\ref{fig:QcrossNLOsg} again for two regions of the invariant mass.
The effect of the NLO correction and the dependence on factorization and
renormalization scale is similar to the one discussed for benchmark
point (p).
\section{\label{sec:concl}Conclusions}
The threshold behaviour of gluino-squark production at hadron colliders
has been investigated. The NLO Green's function has been evaluated for
gluino and squark in the three possible irreducible representations with
dimensions $\bf{3}$, $\bf{6}$ and $\bf{15}$. The logarithmically
enhanced part of the NLO corrections to the "hard kernel" for the
evaluation of the cross section is included in the analysis.
\par
In contrast to the case of gluino pairs the constituent decay rate is generically
comparable or larger than the binding energy, independent of the
choice of the the SUSY parameters. Hence it is only for the exceptional
case of gluino and squark masses being comparable, where both gluino and squark decay rates are suppressed, that pronounced
resonance behaviour might occur. Nevertheless, attractive final state
interaction, in particular for squark and gluino in the triplet
representation, will lead to a modification of the cross section in the
threshold region. The effective threshold is lowered by about
$10\,\mbox{GeV}$ from final state interaction and additional smearing due to the constituent decay rate. Most of the enhancement is concentrated in an
interval of about $50\,\mbox{GeV}$ around the nominal location of the
threshold. Final state interaction, thus, leads to a significant
distortion of the differential distribution
$\mbox{d}\sigma/\mbox{d}M$. Compared to the total cross section evaluated
in NLO without inclusion of the rescattering effects final state
interaction increases the result by about $2$ -- $3\%$.
\begin{appendix}
\section{\label{app:3times8}Tensor product in a general SU(N)}
The tensor product of the adjoint and the fundamental representations of
$SU(N)$ can be obtained following standard methods and is given by
\begin{eqnarray}
(N^2-1)\otimes
N&=&\frac{N(N+2)(N-1)}{2}\oplus\frac{N(N+1)(N-2)}{2}\oplus N
\,,
\label{eqn:NtimesN2-1}
\end{eqnarray}
where the second representation on the right hand side of
Eq.~(\ref{eqn:NtimesN2-1}) is understood to be a conjugate one. (See
also \cite{Kats:2009bv}.)
\par
The projectors for these three representations are given as (for $SU(3)$
see also Eq.~(\ref{eqn:PROJ}))
\begin{eqnarray}
\mathbb{P}^{[R]}_{ai,bj}&=&a^{[R]}\,\delta_{ab}\,\dblone_{ij}+b^{[R]}\,d_{abm}\,T^m_{ij}+c^{[R]}\,i\,f_{abm}\,T^m_{ij}
\,.
\label{eqn:PROJsg}
\end{eqnarray}
The constants $a^{[R]}$, $b^{[R]}$ and $c^{[R]}$ can be determined using
the completeness relation
\begin{eqnarray}
\sum_i\mathbb{P}^{[R_i]}_{aj,bk}&=&\delta_{ab}\,\dblone_{jk}
\,,
\label{eqn:EqSys1}
\end{eqnarray}
the projector properties
\begin{eqnarray}
\mathbb{P}^{[R_1]}_{ai,bj}\mathbb{P}^{[R_2]}_{bj,ck}&=&\mathbb{P}^{[R_1]}_{ai,ck}\,\delta_{R_1R_2}
\,,
\label{eqn:EqSys2}
\end{eqnarray}
as well as the proper dimensionality of the representations
\begin{eqnarray}
\delta_{ab}\,\delta_{ij}\,\mathbb{P}^{[R]}_{bj,ai}&=&d_R
\,.
\label{eqn:EqSys3}
\end{eqnarray}
The result is given in Tab.~\ref{tab:APPprojSG} and the values for $N=3$
coincide with those of Tab.~\ref{tab:PROJ}.
\par
The coefficients governing the strength of the potential, $C^{[R]}$, are
easily obtained by projecting the amplitude for the one gluon exchange
$T^a_{ij}F^a_{bc}$ onto the three different representations
\begin{eqnarray}
d_RC^{[R]}&=&\mathbb{P}^{[R]}_{dh,bi}\,T^a_{ij}F^a_{bc}\,\mathbb{P}^{[R]}_{cj,dh}
\,,
\end{eqnarray}
and the result is listed in Tab.~\ref{tab:APPprojSG}.
\begin{table}[t]
\begin{center}
\begin{tabular}{c||c|c|c||c}
$R$&$a^{[R]}$&$b^{[R]}$&$c^{[R]}$&$C^{[R]}$
\\\hline\hline
$N$&$\frac{1}{N^2-1}$&$\frac{N}{N^2-1}$&$\frac{N}{N^2-1}$&$\hspace{0.4cm}\frac{N}{2}\hspace{0.4cm}$
\\\hline
$\frac{N(N+1)(N-2)}{2}$&$\frac{N-2}{2(N-1)}$&$-\frac{N}{2(N-1)}$&$\frac{N-2}{2(N-1)}$&$\frac{1}{2}$
\\\hline
$\hspace{0.2cm}\frac{N(N+2)(N-1)}{2}\hspace{0.2cm}$&$\hspace{0.2cm}\frac{N+2}{2(N+1)}\hspace{0.2cm}$&$\hspace{0.2cm}\frac{N}{2(N+1)}\hspace{0.2cm}$&$\hspace{0.2cm}-\frac{N+2}{2(N+1)}\hspace{0.2cm}$&$-\frac{1}{2}$
\end{tabular}
\caption{Coefficients of the projectors from Eq.~(\ref{eqn:PROJsg}) for
  a general $SU(N)$.}
\label{tab:APPprojSG}
\end{center}
\end{table}
\end{appendix}


\end{document}